\documentstyle[aps,preprint]{revtex}
\tightenlines

\begin{document}
\title{Experimental study of Taylor's hypothesis 
in a turbulent soap film}

\author{Andrew Belmonte\thanks{Present address: Department of 
Mathematics, Pennsylvania State University, University Park, PA 16802, 
USA.}, Brian Martin, and Walter I. Goldburg}

\address{Department of Physics and Astronomy,
University of Pittsburgh, Pittsburgh, PA 15260, U.~S.~A.\\}

\date{\today} 

\author{\parbox{430pt}{\vglue 0.3 cm \small An experimental study of 
Taylor's hypothesis in a quasi-two-dimensional turbulent soap film is 
presented.  A two probe laser Doppler velocimeter enables a 
non-intrusive simultaneous measurement of the velocity at spatially 
separated points.  The breakdown of Taylor's hypothesis is quantified 
using the cross correlation between two points displaced in both 
space and time; correlation is better than 90\% for scales less 
than the integral scale.  A quantitative study of the decorrelation 
beyond the integral scale is presented, including an analysis of the 
failure of Taylor's hypothesis using techniques from predictability 
studies of turbulent flows.  Our results are compared with similar 
studies of 3D turbulence.}}

\author{\parbox{430pt}{\vglue 0.3 cm \small PACS Number(s):  
47.27.Gs, 68.15.+e, 03.40.Gc}}

\author{{\small REVISED VERSION}}

\maketitle

\narrowtext

\section {Introduction}

In a 1938 paper on the statistics of turbulence, G. I. Taylor 
presented an assumption from which he could infer the spatial 
structure of a turbulent velocity field from a single point 
measurement of its temporal fluctuation \cite{taylor38}.  This 
assumption, known as Taylor's hypothesis or the frozen turbulence 
assumption, relies on the existence of a large mean flow which 
translates the fluctuations past the stationary probe in a time short 
compared to the evolution time of the turbulence.  The experimental 
measurements treated by Taylor were made on the turbulence generated 
behind a stationary grid in a wind tunnel, and his hypothesis has 
become a standard technique employed in similar experiments which 
inform our current views on turbulence (see for example 
Refs.~\cite{monyag}, \cite{eurofl}, and \cite{sreeni98}).  The 
importance of this hypothesis stems from the fact that most turbulence 
theories are framed in terms of the spatial structure of the velocity 
field \cite{monyag,frisch}.

In practical terms, the limits of Taylor's hypothesis are determined 
by how large the mean velocity must be relative to the fluctuations.  
Recently Yakhot \cite{yakhot} has pointed out that the corrections to 
Taylor's hypothesis could well be of the same order as corrections to 
the standard model of turbulence, Kolmogorov's 1941 theory 
\cite{k41a,k41c}.  There is at present, however, no firm theoretical 
derivation of the hypothesis, and thus no fashion to evaluate its 
reliability, calculate higher order corrections, or predict in what 
way it will break down.  A few theoretical discussions do exist 
\cite{lin,lum65,tenn75}, and a promising direction has recently been 
taken by Hayot \& Jayaprakash for the Burgers equation \cite{jay}.  
The treatment by Lumley in particular offers corrections to 
statistical measures of spatial gradients for non-negligible velocity 
fluctuations \cite{lum65}.  Still a general theoretical framework is 
lacking.

Experimentally there have been many studies of Taylor's hypothesis, 
all of which have treated three dimensional (3D) turbulence; we give a 
brief overview in Section II. Our experiments, however, are performed 
on the quasi-two dimensional flow of a soap film.  Two-dimensional 
(2D) fluid flows occur in many physical situations, mostly due to the 
effects of rotation or stratification in the atmosphere and ocean 
\cite{salmon}.  Turbulence in 2D is different from 3D in several ways, 
largely due to the absence of vortex stretching in 2D \cite{lesieur}.  
This means that the squared vorticity (called enstrophy) becomes a 
nearly-conserved quantity in 2D, like the energy, and thus two 
cascades are expected: a {\it direct} cascade of enstrophy to smaller 
scales, and an {\it inverse} cascade of energy to larger scales 
\cite{kraich67}; for decaying 2D turbulence the inverse cascade is 
apparently absent \cite{lesieur,batch69}.  Although Taylor's 
hypothesis is also an important assumption in the study of 2D 
turbulence, to our knowledge it has never been tested, nor is it 
clear that the hypothesis should do relatively better or worse than 
in 3D.

It may be helpful at this point to discuss the essential differences 
between 2D and 3D turbulence in order to better relate the present 
measurements to prior three dimensional tests of the Taylor 
hypothesis.  In three dimensions the vorticity vector ${\bf 
\omega}(x,y,z,t)$ can point in any direction, but in 2D it is 
restricted to be perpendicular to the $x,y$ plane of the flow.  This 
fact alone assures that in inviscid flows, the enstrophy $\Omega 
={\frac{1}{2}} <\omega^{2}>$ is a constant of the motion, in addition 
to kinetic energy conservation $K={\frac{1}{2}} <v^{2}>$ (the angular 
brackets designate an appropriate average).  From the Navier-Stokes 
equation it follows that in 2D, vorticity cannot be amplified (or 
attenuated except by viscous damping) by a velocity field gradient.  The existence of vortex 
stretching in 3D is intimately related to the energy cascade from 
large scales to small, with the process controlled by the rate 
$\epsilon$ at which kinetic energy is injected at large scales.

In 2D, one expects that energy injected at intermediate scale, 
$r_{inj}$ will be transferred to larger scales 
\cite{lesieur,kraichmont} and dissipated at the boundaries of the 
system.  This inverse cascade process is not expected to be local in 
k-space.  For scales $r < r_{inj}$, the small-scale velocity 
fluctuations $<|\delta v(r)|>$, which carry little of the energy, are 
expected to cascade down to the dissipative scale under the control of 
the enstrophy injection rate $\beta \equiv \partial \Omega/\partial t$ 
(the energy injection rate $\epsilon$ plays the corresponding role in 
the 2D inverse cascade).

It seems reasonable that the absence of vortex stretching in 2D will 
increase the likelihood of a velocity fluctuation being transported 
intact to a distant downstream point, thereby enhancing the validity 
of Taylor's hypothesis over 3D turbulence.  Taylor's hypothesis can 
also fail when a local fluctuation is transported laterally into the 
flow path between adjacent points separated by the distance $\Delta x 
= U_{0}\Delta t$.  For purely geometrical reasons, this would seem to 
be a less likely occurence in 2D than in 3D, so that the frozen 
turbulence relation $\delta v(\Delta x) = \delta v(U_{0}t)$ should 
presumably hold out to larger values of $\Delta x$.  It is harder for 
us to assess the impact of 2D nonlocality on the validity of the 
Taylor hypothesis in 2D as compared to 3D.

Our experiment uses a laser Doppler velocimeter (LDV) with two probes 
to nonintrusively examine Taylor's hypothesis in a turbulent soap 
film.  We differentiate between two different aspects of the 
hypothesis.  By {\it Taylor's hypothesis of coherence} we mean the 
assertion that the velocity field is unchanged as it is advected 
downstream: that $v(x,t)$ is identical to $v(x+\Delta x,t+\tau_{0})$, 
where $\tau_{0} = \Delta x/U_{0}$, with $U_{0}$ being the mean 
velocity and $\Delta x$ the distance between the two points 
\cite{taylor38}.  For a clear illustration of this, glance ahead to 
Figure \ref{f-vt}.  This coherence hypothesis is unquestionably an 
approximation and must fail as $\Delta x$ becomes large; here we 
quantify this failure, and relate it to the predictability problem for 
turbulent flows.  We measure the breakdown of the coherence hypothesis 
via the correlation between two points in the flow, displaced in both 
position and time 
\cite{favre57,favre58,fisher64,champagne,comte71,cenedese91}.  We find 
nevertheless that the expectation value of the lowest six moments of 
the longitudinal velocity difference is unaffected by the loss of 
coherence.  Thus time correlation statistics of velocity fluctuations 
appear to be the same as spatial ones; we will refer to this as {\it 
Taylor's statistical hypothesis}, which is implied by the coherence 
hypothesis, but does not actually require it.

\section {Previous Studies of Taylor's Hypothesis}

Taylor's hypothesis has been exposed to many experimental tests in 
three dimensional turbulence, and we do not intend to give an 
exhaustive survey here.  These experimental studies can be divided 
into two broad categories, concerned with either correlations over 
finite distances 
\cite{favre57,favre58,fisher64,champagne,comte71,cenedese91}, or local 
spatial derivatives used in turbulent dissipation estimates 
\cite{pio89,kail93,mi94,dahm97}.  It has long been appreciated that 
the validity of Taylor's frozen turbulence assumption requires the 
smallness of the turbulent intensity $I_{t}$, defined as the ratio of 
rms velocity fluctuations to the mean flow speed $U_{0}$.  
Additionally, the mean shear rate and the viscous damping must be 
small in the range of spatial scales $\Delta x$ being probed.  In this 
section we give a sampling of the sort of work which has been done in 
3D; a more detailed discussion of studies with which we directly 
compare our results is given in Section V. The reader is referred to 
introductory reviews in two recent articles \cite{mi94,dahm97}, which 
are somewhat complementary to what is given here.

To test the application of Taylor's hypothesis to velocity 
correlations over finite distances, one approach is to compare 
measurements made at a single observation point with measurements made 
at points displaced downstream.  The first such tests were made in a 
wind tunnel by Favre, Gaviglio, \& Dumas \cite{favre57,favre58}.  The 
measurements were performed within the turbulent boundary layer of a 
plate at various $I_{t}$ up to 15\%.  They found that Taylor's
statistical hypothesis is valid for measurements of the velocity 
correlation function $R(\Delta x, \tau) = \langle v_{1}(x_{1},t) 
v_{2}(x_{1}+\Delta x,t - \tau) \rangle$ made not too close to the 
plate.  Fisher \& Davies \cite{fisher64} made careful measurements of 
the velocity correlation function $R(\Delta x, \tau)$ in a jet. They 
found that the relation $\tau =\Delta x/U_{0}$ was not well satisfied 
in the mixing region, where $I_{t}$ is typically $\sim$ 20\%.  These 
authors observed, as have many others, that the functional form of $R$ 
changes with increasing $\Delta x$, and that this function is not very 
sharply peaked, as it would be if Taylor's coherence hypothesis were 
satisfied.  Comte-Bellot \& Corrsin \cite{comte71} measured $R(\Delta 
x, \tau)$ for grid-generated turbulence in a wind tunnel, where both 
$I_{t}$ and the mean shear rates are rather small.  Though downstream 
decay causes the maximum value of $R$ vs.~$\tau$ to decrease with 
increasing $\Delta x$, the correlation functions could still be 
collapsed onto the same functional form.

One of the fundamental effects of turbulence is its enhancement of 
dissipation, and measurements of this require knowledge of spatial 
derivatives.  Using Taylor's hypothesis allows the time deriative of a 
single point measurement to be related to spatial derivatives, the 
simplest relation being $\partial \phi/\partial x =U_{0}\partial 
\phi/\partial t$, where $\phi$ could be a passive scalar 
concentration, temperature, a velocity component, or a product of 
velocity components.  Kailasneth, Sreenivasan, \& Saylor \cite{kail93} 
studied a variety of turbulent systems and tested Taylor's hypothesis 
using a fluorescent dye in a jet, the heated wake of a cylinder, and 
the atmospheric boundary layer.  They were interested in the 
probability density function of these scalars, and found that Taylor's 
statistical hypothesis worked well for conditional probability 
densities of the scalar fluctuations.  Mi \& Antonia \cite{mi94} used 
a heated jet of air, with a turbulent intensity of about 26 \%, to
verify different theoretical relations between spatial and temporal 
derivatives of temperature, corrected for finite turbulent 
intensities.  Dahm \& Southerland \cite{dahm97} compared 2D spatial 
and spatio-temporal gradient fields of fluorescent dye in a water jet 
using a fast photodiode array.  They found that Taylor's coherence 
hypothesis was only approximately verified for these fields.  
Piomelli, Balint, \& Wallace \cite{pio89} studied Taylor's hypothesis 
for various velocity derivatives and compared hot wire measurements, 
large eddy simulations, and direct numerical simulations of the 
Navier-Stokes equation for wall-bounded flows.  Taylor's hypothesis 
was found to be in accord with the calculations and measurements made 
sufficiently far from the wall, where the mean shear is not excessive.  
In our experiments we have not treated the application of Taylor's 
hypothesis to gradients.

All hot wire measurements are sufficiently intrusive that one must 
compensate or otherwise adjust for the perturbations produced by the 
wake of the upstream probe on the velocity measured at the downstream 
probe.  Cenedese {\it et al.} \cite{cenedese91} avoided this problem 
by making velocity measurements with a Laser Doppler system.  Only for 
small $\Delta x$ did their correlation measurements satisfy Taylor's 
hypothesis very well, though admittedly $I_{t}$ was rather high 
(13\%).  We discuss their results in more detail in Section
V.

In the present experiment, Laser Doppler velocimetry is also used to 
measure $R(\Delta x, \tau)$, so that there is no perturbation on the 
downstream probe.  In contrast to the studies discussed above, 
however, we examine Taylor's hypothesis in a quasi-2D system, a 
flowing soap film; we have also studied the effects on the higher 
order moments of velocity differences.  Before presenting our results, 
we discuss in some detail our experimental system.

\section {Experimental Setup}

The use of soap films as convenient systems for the experimental study 
of 2D hydrodynamics began with the pioneering work of Couder and 
coworkers \cite{coud81,coud84,chomaz} and Gharib and Derango 
\cite{ghar89}.  Our measurements are performed using a flowing soap 
film apparatus developed at the University of Pittsburgh by Kellay, 
Wu, \& Goldburg \cite{kellay}, and Rutgers, Wu, \& Goldburg 
\cite{maasetup,maa97}; we are using the latest version of this system, 
built by Rutgers.  In our setup, a thin soap film several $\mu$m thick 
is allowed to fall between two taut plastic wires from an upper 
reservoir into a lower one, see Figure \ref{f-setup}.  The channel 
width is $W = 6.2$ cm over a distance of 120 cm, where the 
measurements are performed.  Quasi-two-dimensional turbulence is 
generated behind a comb (tooth diameter 1 mm and spacing $M =$ 3.8 mm) 
which perforates the film at a fixed height.  The typical transit time 
between comb and lower reservoir is the order of a second.

The turbulence generated in such a soap film decays downstream from 
the grid, exhibiting many aspects which agree with theories of 2D 
turbulence \cite{coud84,chomaz,ghar89,kellay,bmartin98,maa98}, though 
there are also some differences.  It therefore seems worthwhile to 
evaluate the experimental situation to date.  Because the lateral 
dimensions of soap films are many orders of magnitude larger than 
their thickness, there would seem to be no doubt that the vorticity 
can indeed be regarded as a scalar quantity, so that vortex stretching 
is absent \cite{frisch}.  This would be a key requirement for the 
occurence two-dimensional turbulence in soap films.  In addition, to 
be able to compare with theoretical and numerical results concerning 
incompressible turbulence, the two-dimensional compressibility of the 
film should be zero.  This condition is clearly not fully met, since 
fluctuations in film {\it thickness} are visible to the eye through 
optical interference of light reflected from the front and back faces 
of the film \cite{chomaz,ghar89,maa98}.  There are, however, recent 
experiments in which the two dimensional divergence $D_{2}={\bf 
\nabla}_{2} \dot {\bf v}(x,y)$ was measured by particle imaging 
velocimetery \cite{riv98}.  In a set-up very similar to that used 
here, $D_{2}$ was measured to be 10-15\% of the rms vorticity near the
comb.  Note also that because the velocity of peristaltic waves in a 
soap film is orders of magnitude larger than the turbulent velocity 
fluctuations \cite{chomaz}, it is expected that the film may be 
regarded as incompressible from the point of view of this study.

Another mitigating factor to the two-dimensionality of soap film flow 
is the friction between the film and the surrounding air.  In a set of 
experiments in which the film was placed in a partial vacuum where the 
air pressure 3 \% of the
atmospheric value, it was found that the energy spectrum $E(k)$ 
decayed for a decade in $k$ as $E(k) \propto k^{-\zeta}$, with $\zeta 
= 3.3 \pm$ 0.3.  This exponent had the same value in both the partial 
vacuum and at atmospheric pressure.  The main effect of the reduced 
pressure was to magnify the magnitude of $E(k)$ near the comb 
\cite{bmartin98}.  These measurements also showed that the total 
kinetic energy initially decayed downstream, but ultimately leveled 
off at a sufficient distance below the comb, which is expected 
theoretically at very high Reynolds numbers \cite{batch69}.  If the 
levelling off distance is called $x^{*}$ and the corresponding time 
$t^{*} =x^{*}/U_{0}$, then all measurements reported in this paper 
were made at values of $t < t^{*}$, i.e.~at distances from the comb 
too close for the levelling off to have occurred.

In most (but not all) measurements of decaying turbulence in a soap 
film, only the enstrophy cascade is observed, i.e.  the inverse energy 
cascade ($E(k) \sim k^{-5/3}$) is not.  Recently an experiment was 
performed in which turbulence was forced by an array of teeth parallel 
to the direction of flow \cite{maa98}.  There one finds evidence of 
both the inverse cascades ($k^{-5/3}$) and the inverse cascade 
spectrum, where $k^{-3}$.

We use a commercial LDV system \cite{TSI} to measure the film velocity 
fluctuations \cite{melling,buchave}.  The soap solution (water and 2\% 
commercial detergent by volume) is seeded with 1 $\mu$m polystyrene 
spheres at a volume fraction of about $10^{-4}$, and the data rate 
ranges from 1 to 8 kHz.  At a distance $x =$ 8 cm below the comb, the 
mean and RMS velocities were typically $U_{0} =$ 180 cm/s and $v_{RMS} 
\equiv \langle v'^{2}\rangle^{1/2} =$ 24 cm/s in the longitudinal 
(streamwise) direction, where $v' \equiv v - U_{0}$.  The turbulent 
intensity in these experiments was $I_{t} \simeq 0.14$.  This is the 
quantity which is assumed in Taylor's hypothesis to be small 
\cite{taylor38,lin}, and we have explicitly chosen for our study a 
value of $I_{t}$ which is not very small.  The Reynolds number for the 
channel is $Re_{W} \equiv U_{0}W/\nu \simeq 11,000$, and for the comb 
$Re_{M} \equiv U_{0}M/\nu \simeq 700$.  The viscosity of a flowing 
soap film is not a well established quantity; here we use $\nu = 
0.1$~cm$^{2}$/s as measured in a 2D Couette viscometer by Martin \& Wu 
\cite{mart95}.  Deviations from two-dimensionality caused by air 
friction \cite{maa97} appear not to affect the turbulence 
for the scales of interest here \cite{bmartin98}.

In order to test Taylor's hypothesis, two LDV heads are used at 
spatially separated points.  Figure 1 shows the arrangement, with the 
downward (flow) direction defined as the $x$ direction.  One head 
(labeled ``LDV 1'') is kept fixed at $x_{1} =$ 8 cm below the comb, 
while a second head (``LDV 2'') is placed at various points ranging 
from $x_{2}$ = 4 to 30 cm below the comb.  The two probes are arranged 
to be directly in line with each other, so that the lower probe 
measures the same part of the flow as the upper one, with a delay 
given by the transit time between them.

Because the LDV only measures velocity when there is a scatterer in 
its measuring volume, the two probes do not in general measure 
velocity simultaneously.  Therefore some binning or ``simultaneity 
window'' $\Delta t$ is needed to perform statistical comparisons: two 
measurements are treated as simultaneous if they occur within $\Delta 
t$ of one another.  Here we use binning windows from $\Delta t =$ 25 
$\mu$s to 200 $\mu$s, corresponding to frequencies $1/\Delta t$ from 5 
to 40 kHz, which are higher than the largest observed frequency in the 
velocity power spectrum.  All measurements reported here are 
insensitive to small changes in $\Delta t$.

One of the difficulties in examining the validity of Taylor's 
hypothesis in decaying turbulence is that the ``small parameter'' 
$I_{t}$ is not constant, but decreases downstream as the turbulence 
decays.  In Fig.~\ref{f-It}a we plot our turbulent intensity as a 
function of distance from the comb.  Thus, although we measure the 
correlation of velocity fluctuations relative to $x_{1} = 8$ cm, where 
$I_{t} \simeq$ 0.14, the actual turbulent intensity affecting the 
velocity field downstream is always less than 0.14.  This would seem 
to significantly complicate the matter.

How does the turbulence decay in our system?  A standard result from 
3D decaying grid turbulence is that the inverse square of the 
turbulent intensity depends on the distance from the grid as: 
$I_{t}^{-2} = A(x/M - B)^{\beta}$, where the dimensionless constants 
are typically found to be $A \sim$ 130 - 150 and $B \sim$ 3 - 20 for 
$\beta = 1$ \cite{bt48init}, or $A \sim$ 20 and $B \sim$ 3.5 for 
$\beta = 1.25$ \cite{comte71}; note that $B$ is the effective position 
of the origin for this scaling in units of $M$.  Assuming that $\beta 
= 1.25$ means that $I_{t}^{-1.6}$ should be a linear function of 
$x/M$; however we find that by taking $I_{t}^{-1.1}$ we get the best 
linear plot vs.~$x/M$, as shown in Fig.~\ref{f-It}b.  The line 
corresponds to
$$
\frac{1}{I_{t}^{2}} = 0.2 \left(\frac{x}{M}\right)^{1.8},
$$
with $B$ = 0 for our fit, which means that the virtual origin is 
located at the position of the grid itself.  Note that we do not 
measure far enough downstream \cite{bmartin98} to see any evidence of 
the kinetic energy saturation as the turbulence decays downstream 
($I_{t} \sim$ constant).

\section {Results}

\subsection{Testing Taylor's Coherence Hypothesis}

Figure \ref{f-vt} shows the velocity fluctuations measured by two 
probes with $\Delta x =$ 0.5 cm.  For this small separation, Taylor's 
hypothesis is clearly a good estimate: the two velocity traces are 
nearly identical except for a small shift in time, which should 
correspond to the transit time across the spatial separation $\Delta 
x$.  To test whether the velocity trace is translated spatially 
without evolving dynamically, we measure the cross-correlation 
$C_{12}(\tau,\Delta x,x_{1})$ between the two probes:
\begin{equation}
C_{12}(\tau,\Delta x,x_{1}) \equiv \frac{\langle v_{1}(x_{1},t) 
v_{2}(x_{1}+\Delta x,t-\tau) \rangle}{v_{1 RMS} \times v_{2 RMS}}.
\label{e-C12}
\end{equation}
Here $v_{1}(x_{1},t)$ and $v_{2}(x_{2},t)$ are the two velocities 
measured by the probes, $v_{i RMS}$ are the RMS velocity fluctuations, 
$\Delta x \equiv x_{2} - x_{1}$, and the brackets $\langle \rangle$ 
denote a time average.  Note that $C_{12}$ is a function not only of 
delay time $\tau$ and separation $\Delta x$, but of the absolute 
location of the first probe $x_{1}$.  This last dependence comes from 
the fact that the turbulence is decaying.  In this study we fix $x_{1} 
=$ 8 cm, and ignore this dependence.

Figure \ref{f-C12} shows $C_{12}(\tau,\Delta x)$ vs.  $\tau$ for several 
different separations $\Delta x$.  As expected, there is a well-defined 
maximum correlation $$C_{12}^{MAX}(\Delta x) \equiv C_{12}(\tau_{MAX},\Delta x)$$ 
at a particular value of the delay time $\tau_{MAX}(\Delta x)$.  
Taylor's coherence hypothesis requires that $C_{12}^{MAX}(\Delta x)$ 
be close to 1, and $\tau_{MAX}(\Delta x)$ be equal to the transit time 
$\Delta x/U_{0}$.  Figure \ref{f-tt} shows $\tau_{MAX}$ as a function 
of $\Delta x/U_{0}$, in agreement with the line drawn for 
$\tau_{MAX}=\Delta x/U_{0}$ \cite{Ucalc}.  As predicted by Taylor's 
hypothesis, the slope of this line is unity.  The small deviations are 
due to errors in our measurement of $\Delta x$.

In Figure \ref{f-C12MAX} we plot the maximum correlation 
$C_{12}^{MAX}(\Delta x)$.  As one would expect, the correlation 
decreases as $\Delta x$ increases, though we have not found any simple 
functional form to fit to this decrease, nor is there to our knowledge 
any predicted form.  The loss of correlation is due to the dynamic 
evolution of the velocity fluctuations and sets a limit to 
Taylor's hypothesis, which we quantify by defining an ``evolution 
length'' $\delta_e$ as the separation for which the correlation drops to 
50\%.  For our experiment
we find $\delta_e \simeq$ 7 cm, corresponding to an evolution time 
$\tau_{e} = \delta_e/U_{0} \simeq 40$ ms.  This length is much larger 
than the relevant lengths of the turbulent velocity field, as we show 
next.

\subsection{Testing Taylor's Statistical Hypothesis}

The statistical study of turbulence is framed in terms of velocity 
correlation functions, structure functions, and energy spectra; here 
we focus on the structure function.  The longitudinal velocity 
difference between two points separated by a distance $r$ is written 
as
$$
\delta v (r,t) 
\equiv ({\bf v(}x_{1} + r,t) - {\bf v(}x_{1},t))\cdot {\bf \hat{r}},$$ 
where the unit vector ${\bf \hat{r}}$ is in the downward 
direction of the flow.  The $n$th order structure function is defined as:
$$
S_{n}(r) \equiv \langle ( \delta v(r,t))^{n} \rangle.
$$
In Figure \ref{f-mom} we show the structure functions $S_{2}(r)$, 
$S_{4}(r)$, and $S_{6}(r)$, calculated from single point velocity 
measurements using Taylor's hypothesis: $r = U_{0} \tau$ (solid 
circles).  We also plot direct spatial measurements of these structure 
functions made using two probes (open squares).  This is a direct 
confirmation of Taylor's statistical hypothesis, which is one of the 
central results of our study.  Note that $S_{2}(r)$ shows a scaling 
region of about a decade where $S_{2}(r) \propto r^{1.6}$, in good 
agreement with other experiments on turbulent soap films 
\cite{belm98,riv98}.  We also find approximately that $S_{4}(r) 
\propto r^{2.9}$ and $S_{6}(r) \propto r^{4.0}$, as shown in the 
figure.  The third moment $S_{3}(r)$ has been treated in detail 
elsewhere \cite{belm98}.

Some comment must be made on the observed values of the exponents, 
which are so different from $S_{n}(r) \propto r^{n}$, the theoretical 
expectation for the enstrophy cascade in 2D turbulence 
\cite{batch69,kraich67}.  It is well known that in 3D turbulence the 
scaling law exponents of the $n$th order structure functions deviate 
from their expected value of $n/3$ as $n$ gets large \cite{ansel84}.  
This systematic difference is attributed to the intermittency of the 
fluctuations \cite{frisch}.  However, the third order structure 
function must scale as $S_{3}(r) \sim r$ even with intermittency, as 
can be derived directly from the Navier-Stokes equations 
\cite{frisch}.  An equivalent derivation for 2D turbulence would not 
be relevant to the third moment in the enstrophy cascade range 
discussed here.  In fact $S_{3}$ is observed to be approximately zero 
in this range (and is positive for large $r$) \cite{belm98}.  This 
observation suggests to us that the scaling exponents of the enstrophy 
range are more sensitive to intermittency than in 3D turbulence.  
Evidence of intermittency, indicated by non-Gaussian velocity 
fluctuations, has been reported previously for our experiment 
\cite{belm98}.

For $r >$ 1 cm, the $S_{n}(r)$ saturate to constant values, which for 
$S_{2}(r)$ is equal to $2v_{RMS}^{2}$.  This occurs roughly at the 
integral or outer scale \cite{monyag}, which characterizes the largest 
scales on which the velocity is correlated.  The integral scale 
$\ell_{0}$ is defined as
\begin{equation}
\ell_{0} \equiv \int^{\infty}_{0}{b(r)dr} /v^{2}_{RMS},
\label{e-intscale}
\end{equation} 
where $b(r)$ is the velocity correlation function:
$$
b(r) \equiv \langle v'(x,t)v'(x+r,t) \rangle = v^{2}_{RMS} 
- \frac{1}{2} S_{2}(r). 
$$
At $x_{1} =$ 8 cm we find $\ell_{0}=$ 0.6 cm, which is much less than 
the evolution length $\delta_e$ = 7 cm.  Thus for the turbulence in 
our soap film, Taylor's hypothesis is justified: the two signals are 
correlated better than 90\% for scales $r < \ell_{0}$ (see 
Figs.~\ref{f-C12MAX} and \ref{f-mom}).

\subsection{Detailed Study of the Velocity Decorrelation}

There are in general two reasons for the breakdown of Taylor's 
hypothesis: the entrance of new structures into the line of travel, 
introducing new fluctuations into the signal, and the evolution of the 
velocity field itself.  In Figure \ref{f-Voverlay} we show an overlay 
of the velocity measured at $x_{1} =$ 8.0 cm vs.~$t$, and the velocity 
measured at $x_{2} =$ 10.0 cm vs.~$t - \tau_{MAX}$ ($\Delta x = 2$ 
cm).  For a perfect correlation ($C_{12}^{MAX} = 1$) the two curves 
would fall on top of each other.  The arrows indicate fluctuations 
which have either appeared or disappeared during the transit time 
between the two probes.  In effect this means that information is 
being generated, and this ``new information'' is partially responsible 
for the velocity decorrelation (Fig.~5).

To explore the details of this process, we measure the coherence 
spectrum $C\!s(f)$ of the fluctuations at $x_{1}$ and $x_{2}$ 
\cite{cospec}.  If Taylor's hypothesis of coherence were justified, 
then $v_{1} \equiv v(x_{1},t)$ would be identical to $v_{2} \equiv 
v(x_{2},t - \tau_{MAX})$.  If $\hat{v}_{1}(f)$ is the complex Fourier 
transform of $v_{1}$, then the standard power spectrum is $P\!s_{1}(f) 
= \langle \hat{v}_{1}(f)\hat{v}^{*}_{1}(f) \rangle$, where 
$\hat{v}^{*}$ is the complex conjugate of $\hat{v}$.  The coherence 
spectrum is
\begin{equation}
C\!s(f) \equiv \frac{1}{2} \frac{\langle \hat{v}_{1}(f)\hat{v}^{*}_{2}(f) 
+ \hat{v}^{*}_{1}(f)\hat{v}_{2}(f)\rangle}{\sqrt{P\!s_{1}(f) \times P\!s_{2}(f)}},
\label{e-cs}
\end{equation}
normalized so that $C\!s =$ 1.0 for frequencies where the two time 
series are coherent.  A measurement of $C\!s(f)$ with increasing probe 
separation shows which modes in the turbulent spectrum persist longer 
and which evolve faster.

Consider first the velocity power spectrum at a single point, shown as 
the thin line in Figure \ref{f-CS}.  This spectrum, according to the 
standard picture of 2D decaying turbulence \cite{kraich67,batch69}, 
should have a power law dependence $P\!s(f) \sim f^{-\alpha}$ in the 
enstrophy cascade range, with $\alpha =$ 3.  In an earlier soap film 
experiment \cite{bmartin98}, this exponent was found to be measurably 
larger than 3; here in we find $\alpha = 3.6 \pm 0.2$.  We compare 
this to the coherence spectrum, which we expect to be nearly equal to 
1.0 for small separations.  The coherence spectrum for $\Delta x =$ 
0.2 cm is also shown in Figure \ref{f-CS} (thick line).  We see that 
$C\!s(f)$ is indeed close to unity over most of the frequency range in 
which the power spectrum appears.  However, the coherence drops below 
$C\!s \sim 0.9$ at $f \sim$ 300 Hz, around the middle of the range 
where $P\!s(f) \sim f^{-\alpha}$, and for $f \sim$ 850 Hz, where the 
power spectrum is reaching the noise floor in our measurement, $C\!s 
\sim 0.2$.  Thus already at $\Delta x =$ 0.2 cm it appears that the 
high frequency components are the most rapidly evolving.  In contrast, 
for the 2D enstrophy cascade it is expected that the `eddy turnover 
time' is independent of size \cite{lesieur}; the observed falloff in 
$C\!s(f)$ may indicate viscous dissipation effects.

The coherence spectra at five increasing separations $\Delta x$ are 
shown in Figure \ref{f-4CS}, as a linear-log plot.  In each case we 
see decorrelation at higher frequencies (smaller scales), while the 
low frequency part remains at a constant value which decreases as 
$\Delta x$ increases.  This constant correlation is approximately 
equal to $C_{12}^{MAX}(\Delta x)$, which means that the overall 
coherence of the velocity field is determined mainly by the low 
frequency components.  The high frequency decorrelation also moves to 
lower frequencies as $\Delta x$ increases.  To see whether the whole 
shape of the $C\!s(f)$ follows the decay of $C_{12}^{MAX}(\Delta x)$, 
we normalize the coherence spectra as $C\!s(f)/C_{12}^{MAX}$ in Figure 
\ref{f-5CSnorm}.  The curves do not lie entirely on top of each other, 
indicating that the cutoff at high frequencies follows a different 
evolution than $C_{12}^{MAX}$.  The cutoff shape is well described as 
$C\!s(f) \sim log(1/f)$, shown as the straight lines drawn through the 
data.  This advancing cutoff is reminescent of the loss of 
predictability in the spectra of atmospheric turbulence simulations 
\cite{lesieur,lesieurerr}.

At larger separations ($\Delta x > 12$ cm), we find that the spectral 
position of this cutoff no longer moves to lower frequencies as 
$\Delta x$ increases.  In Figure \ref{f-L3CS} we superpose the 
coherence spectra for $\Delta x$ from 12 to 22 cm, normalized by 
$C_{12}^{MAX}(\Delta x)$.  The curves lie reasonably on top of each 
other, which means that the entire coherence spectrum follows the 
overall decrease of $C_{12}^{MAX}$.  By taking the intersection of the 
logarithmic fit with the line $C\!s(f) = C_{12}^{MAX}(\Delta x)$ in 
Figs.~\ref{f-5CSnorm} and \ref{f-L3CS}, we use the frequency $f_{d}$ 
of the intersection to characterize the position of the spectral 
cutoff \cite{lesieur}.  We plot this frequency in Figure \ref{f-fd}.  
Up to $\Delta x \simeq 12$ cm, the advance of $f_{d}$ to lower 
frequencies is consistent with the scaling $f_{d} \sim \Delta 
x^{-1/2}$, which is slower than that seen for wavenumber cutoff in 2D 
numerical simulations of atmospheric predictability 
\cite{lesieur,lesieurerr}.  For $\Delta x > 12$ cm, $f_{d}$ reaches a 
constant value of about 30 Hz, corresponding to a length of 6 cm, 
which is the size of our system (the channel width $W = 6$ cm).

\subsection{Turbulent Predictability and Taylor's Hypothesis}

The failure of Taylor's hypothesis in our experiment is closely 
related to the question of predictability in 2D turbulence 
\cite{lesieurerr,leith71,kida90}.  The general study of predictability 
in turbulence (see for example \cite{lesieur,lk72,aurell96}) is of 
particular importance to the weather prediction problem 
\cite{predictref2}.  Here we briefly show how our analysis 
parallels this general framework.  Note that here we are comparing a 
developing turbulent velocity with its initial state, whereas studies 
of predictability treat the diverging evolution of two nearly 
identical initial states.  Nonetheless there are several similarities 
between the two.

Following M\'etais \& Lesieur, we first define the time series of the 
{\it velocity difference}, or error time series \cite{lesieurerr}, 
which for our experiment is written
\begin{equation}
\Delta v(\Delta x,t) \equiv v_{1}(x_{1},t) - v_{2}(x_{1} + 
\Delta x,t - \tau_{MAX}).
\label{e-errdv}
\end{equation}
For a perfectly correlated signal this time series would be 
identically zero.  We define the difference energy $\Delta E(\Delta x) 
= \langle (\Delta v(\Delta x,t))^{2} \rangle$, which is the second 
moment of $\Delta v$ and thus a kinetic energy associated with the 
difference series; it is analogous to the error energy in 
predictability studies \cite{lesieurerr,boffeta}.  In Figure 
\ref{f-rho} we plot a dimensionless $\Delta E$, namely
\begin{equation}
\rho(\Delta x) \equiv \frac{\Delta E(\Delta x)}{v_{1 RMS}^{2} + 
v_{2 RMS}^{2}},
\label{e-rhodef}
\end{equation}
as a function of the decay time $\Delta x/U_{0}$.  The function $\rho$ 
is defined to increase from 0 to 1 as $\Delta x$ increases, and acts 
as a sort of distance function between the two velocities.  By 
comparing Eqs.~\ref{e-C12} and \ref{e-rhodef}, one sees that $\rho$ 
and $C_{12}$ are simply related.

The inset to Fig.~\ref{f-rho} shows an enlargement of $\rho(\Delta 
x/U_{0})$ near $\Delta x/U_{0} = 0$.  There is no clear linear portion 
in the plot which would correspond to an exponential error increase.  
In the study of turbulent predictability an exponential error growth 
is used to define a sort of Lyapunov exponent \cite{eck85}, with the 
error energy serving as a metric for evaluating the distance between 
co-evolving turbulent states.  The data shown in Fig.~\ref{f-rho} are 
in fact better described by a power law $\rho \sim (\Delta 
x/U_{0})^{1/2}$, as shown in Fig.~\ref{f-rho2}.  This apparent 
square root dependence should be interpreted only as a power law 
dependence: the actual value of the exponent depends on the choice of 
metric function, Eq.~\ref{e-rhodef}.  Since an exponential growth of 
the error energy depends on the linearization of an underlying 
equation for $\rho$, Fig.~\ref{f-rho2} may indicate the presence of 
higher order terms, similar to the quadratic saturation term used by 
Lorenz to fit error growth in an iterated map \cite{lorenz}.

The {\it predictability time} $T_{p}$ is a standard measure of the 
time beyond which one can no longer project the state of a turbulent 
system \cite{lesieurerr,aurell96}.  The exact definition of such a 
time is somewhat arbitrary, though it is usually much larger than the 
large scale eddy turnover time for the turbulent flow 
\cite{lesieurerr}.  We can characterize the long time growth of 
$\rho(\tau)$ by quantifying how long the velocity $v_{2}$ remains 
similar to $v_{1}$.  The predictability time used by M\'etais \& 
Lesieur was defined by $\rho(T_{p}) = 0.5$ \cite{lesieurerr}; we 
define an evolution time $T_{e}$ such that $\rho(T_{e}) = 0.5$, and 
find that $T_{e} \simeq$ 25 ms.  This is of the same order as our 
decorrelation time $\tau_{e} \simeq 40$ ms given by $C_{12}^{MAX}$ 
(Fig.~\ref{f-C12MAX}), a result which is not surprising given that the 
two functions are related.  The analogy between the loss of 
predictability and the failure of Taylor's hypothesis is in fact 
rooted in a common cause: the loss of velocity coherence due to 
turbulence.  Whether any implications can be drawn from this 
connection remains to be seen.

\section {Discussion}

\subsection{Comparison with 3D Measurements}

We have measured the breakdown of Taylor's hypothesis for decaying 
turbulence in a flowing soap film and shown that the hypothesis is a 
valid assumption for statistical measurements of the turbulence (the 
structure functions).  How do our measurements compare to similar 
experimental studies of 3D decaying turbulence?  Of the six studies 
which to our knowledge provide information comparable to 
Fig.~\ref{f-C12MAX} 
\cite{favre57,favre58,fisher64,champagne,comte71,cenedese91} we will 
examine three in detail \cite{champagne,comte71,cenedese91}.  Two of 
these studies used hot wire anemometry \cite{champagne,comte71}, and 
thus additional techniques were required to compensate for the wake of 
the upstream probe.  Champagne {\it et al.}~\cite{champagne} used a 
`grid' made of 12 parallel channels (spacing $M' =$ 2.54 cm) in a wind 
tunnel with a mean speed of 12 m/s, and $Re_{M'} =$ 21,000.  
Cross-correlation measurements started at $x_{1} =$ 259 cm, where 
$I_{t} \simeq$ 0.018 and $\ell_{0} \simeq$ 4.2 cm.  Comte-Bellot \& 
Corrsin \cite{comte71} made measurements behind a standard grid ($M =$ 
5.08 cm) in a wind tunnel with $Re_{M} =$ 34,000.  The two hot-wire 
cross correlation measurements were made starting at $x_{1} =$ 210 cm, 
where $I_{t} \simeq$ 0.022 and $\ell_{0} \simeq$ 1.1 cm.  Cenedese 
{\it et al.}~\cite{cenedese91} used a nonintrusive laser Doppler 
anemometer similar to our LDV (see \cite{buchave}), but did not use a 
standard grid; the turbulence was produced by a combination of a 
honeycomb and the channel walls.  Their measurements were made in a 
water channel (height $h =$ 2 cm) starting at $x_{1} =$ 14 cm, where 
$I_{t} \simeq$ 0.13 and $\ell_{0} \simeq$ 1.0 cm.  Their Reynolds 
numbers were also significantly lower: $Re_{h} =$ 4,800.

Are there any differences between Taylor's hypothesis in our 
approximately two-dimensional soap film and in these 3D experiments?  
We address this question by plotting $C_{12}^{MAX}(\Delta x)$ from 
these three studies along with our measurements in Figure 
\ref{f-C12compar}.  The independent variable in this plot is $\Delta 
x$ in units of the integral scale $\ell_{0}$.  One might expect the 
decorrelation to occur more slowly in the soap film due to the absense 
of vortex stretching.  However, as the turbulent intensity in our 
experiment is high ($I_{t} =$ 0.14) compared to the two wind tunnel 
experiments ($I_{t} \sim$ 0.02), our data should be directly compared 
only to that of Cenedese {\it et al.} ($I_{t} =$ 0.13).  In this case 
we see that indeed the correlation in our soap film extends to much 
larger values of $\Delta x/\ell_{0}$ than in their 3D experiment.  
Note that $C_{12}^{MAX}$ from the wind tunnel experiments also extends 
to much larger values of $\Delta x/\ell_{0}$ than the data of Cenedese 
{\it et al}, probably due to the fact that their turbulent intensities 
are much lower.  More definitive conclusions would come from a single 
experiment (in 2D or 3D) which measures $C_{12}^{MAX}(\Delta x)$ for 
several different $I_{t}$.

\subsection{Detailed Shape of the Cross Correlation $C_{12}(\tau)$}

As of yet there is no rigorous underpinning to Taylor's hypothesis 
which would allow for the calculation of higher order corrections to 
velocity correlations, though an intriguing suggestion was implemented 
in \cite{pinton94}.  To provide detailed information for some future 
theory, we focus on the shape of the cross-correlation function 
$C_{12}(\tau,\Delta x)$ around $\tau_{MAX}$.  This shape is by 
definition (Eq.~\ref{e-C12}) the average convolution of a velocity 
fluctuation taken with itself a time $\tau_{MAX}$ later.  In Figure 
\ref{f-gauss} we show as an example $C_{12}(\tau)$ for $\Delta x =$ 4 
cm, along with a Gaussian distribution centered on $\tau_{MAX}$.  We 
find that the shape is always nearly Gaussian, with a slight skewness 
around $\tau_{MAX}$ consistently towards the positive.  The width of 
the Gaussian does not broaden as $\Delta x$ increases, though the 
maximum does decrease as shown in Figure \ref{f-C12overlay}.  Thus the 
development of the cross-correlation cannot be treated as a 
diffusion-like process, for which the width would increase as the 
maximum decreases.  The small positive skewness is also not strongly 
dependent on $\Delta x$.

\section {Conclusion}

In this paper we focused on the breakdown of Taylor's coherence 
hypothesis in a turbulent soap film, a quasi-2D experimental system.  
We have shown that for the lower order moments the statistical 
hypothesis works well, even when the actual cross correlation between 
the two probes is low.  As the relevant length scale of this 
decorrelation is much larger than the integral scale of the turbulence 
($\delta_{e} >> \ell_{0}$), this phenomenon is outside the region 
usually considered by most studies: it is the turbulence beyond the 
scaling range.  Yet this evolution contains untapped information, as 
we have indicated.  The failure of Taylor's hypothesis may thus shed 
light on deeper problems in turbulence.

\section {Acknowledgements}

We would like to thank M. Rivera and X. L. Wu for beneficial and 
insightful discussions, and H. Kellay, M. A. Rutgers, R. Cressman, and 
the referees for helpful comments on the manuscript.  This work was 
supported by NASA and the National Science Foundation.

\pagebreak


\begin{figure}
\begin{center}
\end{center}														  
\caption{A diagram of the experimental setup: A) upper soap reservoir; 
B) hooks which hold plastic wires between which soap film flows; C) 
comb behind which turbulence is generated in the film; D) lower soap 
reservoir.  The two LDV probes which measure the velocity in the film 
are labeled 1 and 2.  Downstream distances are labeled as referred to 
in the text.}
\label{f-setup}															  
\end{figure}															  
\begin{figure}
\begin{center}
\end{center}														  
\caption{The decay of the turbulent intensity $I_{t}$ behind the comb 
in our soap film: a) $I_{t}$ vs.~downstream distance $x$; b) the same 
data plotted as $I_{t}^{-1.1}$ vs.~$x/M$.  The straight line is a 
linear fit (see text).}
\label{f-It}															  
\end{figure}															  
\begin{figure}
\begin{center}
\end{center}														  
\caption{Simultaneous traces of the velocity vs.~time at $x_{1} =$ 8.0 cm 
and $x_{2} =$ 8.5 cm behind the comb.}
\label{f-vt}														  
\end{figure}															  
\begin{figure}
\begin{center}
\end{center}															  
\caption{The cross correlation $C_{12}(\tau,\Delta x)$ vs. delay time $\tau$
for several different values of $\Delta x$ (as labeled).}	  
\label{f-C12}															  
\end{figure}													  

							  
						  
\begin{figure}
\begin{center}
\end{center}									
						  
\caption{The delay time $\tau_{MAX}$ for the maximum of the cross 
correlations in Fig.~\protect \ref{f-C12} vs.~the transit time $\Delta x / 
U_0$. The solid line corresponds to $\tau_{MAX}$ = $\Delta x / U_0$.}
\label{f-tt}										
					  
\end{figure}									

\begin{figure}
\begin{center}
\end{center}															  
\caption{The maximum value of the cross correlations in Fig.~\protect 
\ref{f-C12}, $C_{12}^{MAX}$ vs.~the downstream separation $\Delta x$.  The 
arrow shows the evolution length $\delta_{e} \simeq$ 7 cm (see text).}
\label{f-C12MAX}															  
\end{figure}															  
\begin{figure}
\begin{center}
\end{center}														  
\caption{A diagram of the experimental setup: A) upper soap reservoir; 
B) hooks which hold plastic wires between which soap film flows; C) 
comb behind which turbulence is generated in the film; D) lower soap 
reservoir.  The two LDV probes which measure the velocity in the film 
are labeled 1 and 2.  Downstream distances are labeled as referred to 
in the text.}
\label{f-setup}															  
\end{figure}															  
\begin{figure}
\begin{center}
\end{center}														  
\caption{The decay of the turbulent intensity $I_{t}$ behind the comb 
in our soap film: a) $I_{t}$ vs.~downstream distance $x$; b) the same 
data plotted as $I_{t}^{-1.1}$ vs.~$x/M$.  The straight line is a 
linear fit (see text).}
\label{f-It}															  
\end{figure}															  
\begin{figure}
\begin{center}
\end{center}														  
\caption{Simultaneous traces of the velocity vs.~time at $x_{1} =$ 8.0 cm 
and $x_{2} =$ 8.5 cm behind the comb.}
\label{f-vt}														  
\end{figure}															  
\begin{figure}
\begin{center}
\end{center}															  
\caption{The cross correlation $C_{12}(\tau,\Delta x)$ vs. delay time
$\tau$
for several different values of $\Delta x$ (as labeled).}	  
\label{f-C12}															  
\end{figure}													  

							  
						  
\begin{figure}
\begin{center}
\end{center}									
						  
\caption{The delay time $\tau_{MAX}$ for the maximum of the cross 
correlations in Fig.~\protect \ref{f-C12} vs.~the transit time $\Delta x / 
U_0$. The solid line corresponds to $\tau_{MAX}$ = $\Delta x / U_0$.}
\label{f-tt}										
					  
\end{figure}									

\begin{figure}
\begin{center}
\end{center}															  
\caption{The maximum value of the cross correlations in Fig.~\protect 
\ref{f-C12}, $C_{12}^{MAX}$ vs.~the downstream separation $\Delta x$.  The 
arrow shows the evolution length $\delta_{e} \simeq$ 7 cm (see text).}
\label{f-C12MAX}															  
\end{figure}															  
\begin{figure}
\begin{center}
\end{center}															  
\caption{Experimental check of Taylor's ``statistical hypothesis'': a) 
second order structure function $S_{2}(r)$ taken at $x =$ 8 cm using 
Taylor's hypothesis; b) fourth order structure function $S_{4}(r)$; c) 
sixth order structure function $S_{6}(r)$.  The open squares are 
direct spatial measurements made using two probes, and the lines 
correspond to the fitted scalings described in the text.}
\label{f-mom}															  
\end{figure}															  
\begin{figure}
\begin{center}
\end{center}															  
\caption{Overlay of the velocity measured at $x_{1} =$ 8.0 cm vs.~$t$, 
and the velocity measured at $x_{1} =$ 10.0 cm vs.~$t - \tau_{MAX}$. 
The arrows indicate fluctuation spikes which have either appeared or 
disappeared during the transit between the two probes.}
\label{f-Voverlay}															  
\end{figure}															  
\begin{figure}
\begin{center}
\end{center}														  
\caption{A comparison of the coherence spectrum (Eq.~\protect 
\ref{e-cs}) for $\Delta x = 0.2$ cm, and the power spectrum taken at 
$x_{1} =$ 8 cm.  The line is a fit to the power law $P\!s_{1}(f) \sim 
f^{-\alpha}$ with $\alpha = 3.6 \pm 0.2$.}
\label{f-CS}															  
\end{figure}															  
\begin{figure}
\begin{center}
\end{center}														  
\caption{Coherence spectra for several values of $\Delta x$ as a
linear-log 
plot of $f$.}
\label{f-4CS}															  
\end{figure}															  
\begin{figure}
\begin{center}
\end{center}														  
\caption{Coherence spectra as in Fig.~\protect \ref{f-4CS} normalized 
by the maximum cross correlation $C_{12}^{MAX}$ from Fig.~\protect 
\ref{f-C12MAX}.  The straight lines through the high frequency part of 
the coherence spectra corresponds to a logarithmic decay law $C\!s(f) 
\sim log(1/f)$.}
\label{f-5CSnorm}															  
\end{figure}															  
\begin{figure}
\begin{center}
\end{center}														  
\caption{Coherence spectra for several large values of $\Delta x$ as a 
linear-log plot of $f$, normalized by the maximum cross correlation 
$C_{12}^{MAX}$ from Fig.~\protect \ref{f-C12MAX}.}
\label{f-L3CS}															  
\end{figure}															  
\begin{figure}
\begin{center}
\end{center}														  
\caption{The frequency scale $f_{d}$ characterizing the logarithmic 
decay $C\!s(f) \sim log(1/f)$ (lines in Fig.~\protect \ref{f-5CSnorm})
vs.~the separation $\Delta x$. The straight line represents the scaling
law
$f_{d} \sim \Delta x^{-1/2}$.}
\label{f-fd}															  
\end{figure}															  
\begin{figure}
\begin{center}
\end{center}														  
\caption{The normalized error energy $\rho(\Delta x/U_{0})$ as a 
function of time $\Delta x/U_{0}$ past the reference point $x_{1}$, 
shown as a log-linear plot.  The arrow indicates the evolution time 
$T_{e}$ as defined in the text.  Inset: an expanded view near the 
origin.}
\label{f-rho}															  
\end{figure}															  
\begin{figure}
\begin{center}
\end{center}														  
\caption{A log-log plot of $\rho(\Delta x/U_{0})$ as a function of 
downstream time $\Delta x/U_{0}$.  The straight line corresponds to a 
power law dependence $(\Delta x/U_{0})^{1/2}$ (see the text).}
\label{f-rho2}															  
\end{figure}															  
\begin{figure}
\begin{center}
\end{center}															  
\caption{A replot of $C_{12}^{MAX}$ from Fig.~\protect \ref{f-C12MAX}
as a function of probe separation $\Delta x$ normalized by the 
integral scale $\ell_{0}$.  Also shown for comparison are 3D results 
from Ref.~\protect \cite{champagne}, Ref.~\protect \cite{comte71}, and 
Ref.~\protect \cite{cenedese91}}
\label{f-C12compar}										
					  
\end{figure}

\begin{figure}
\begin{center}
\end{center}															  
\caption{Gaussian fit to the shape of $C_{12}(\tau)$ at $\Delta x =$ 4
cm.}	  
\label{f-gauss}														  
\end{figure}															  
\begin{figure}
\begin{center}
\end{center}															  
\caption{An overlay of several $C_{12}(\tau)$ at various $\Delta x$ (as 
shown), plotted vs.~$\tau - \tau_{MAX}$.}	  
\label{f-C12overlay}														  
\end{figure}															  


\begin{references}
\bibitem{Ucalc}
For these calculations we recalculate $U_{0}$ for each $\Delta x$ as the 
average of $\langle v_{1}(x_{1},t) \rangle$ and $\langle 
v_{2}(x_{2},t) \rangle$.
\end{references}
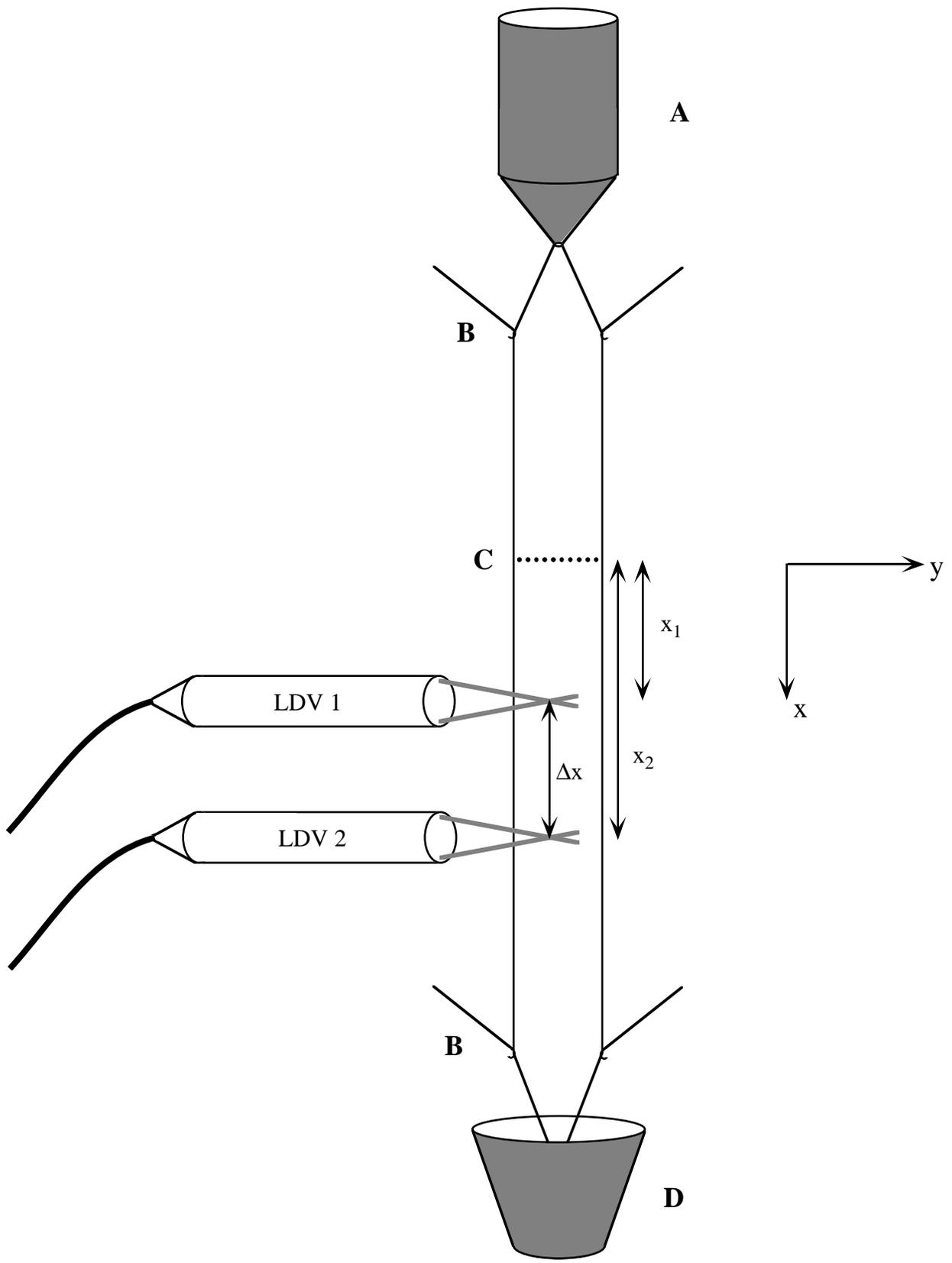
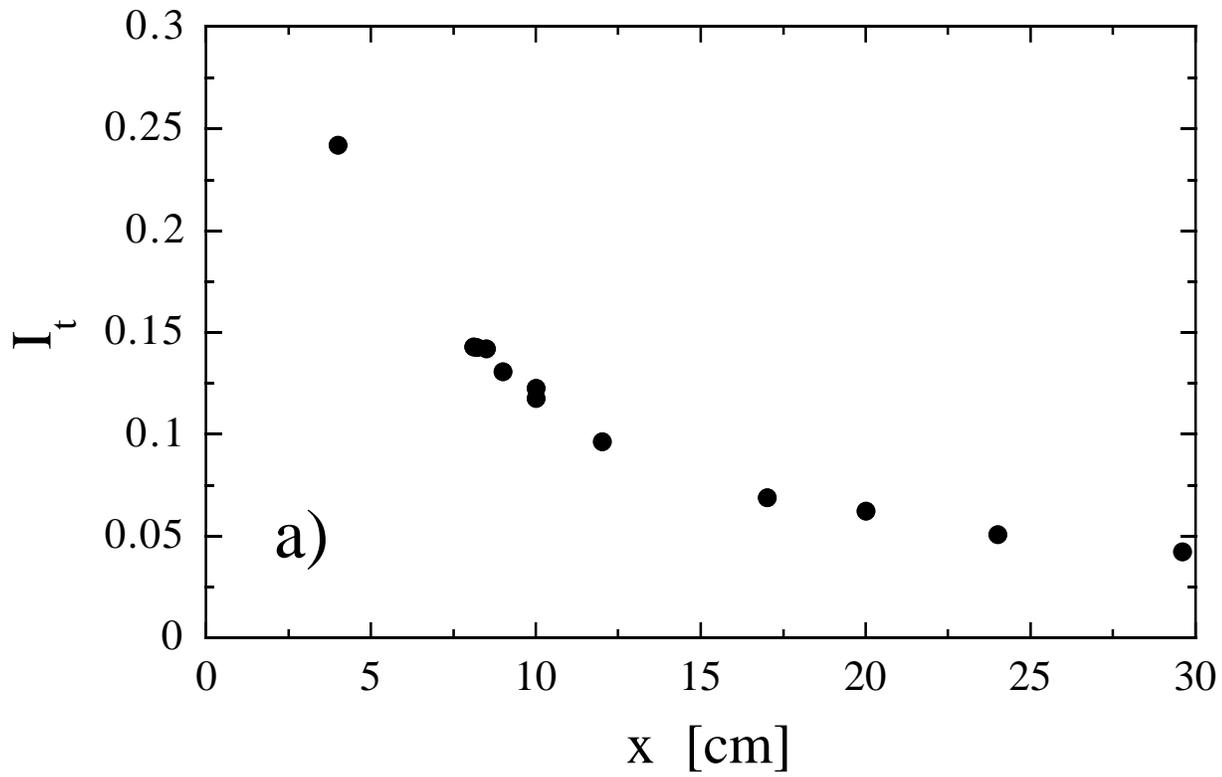
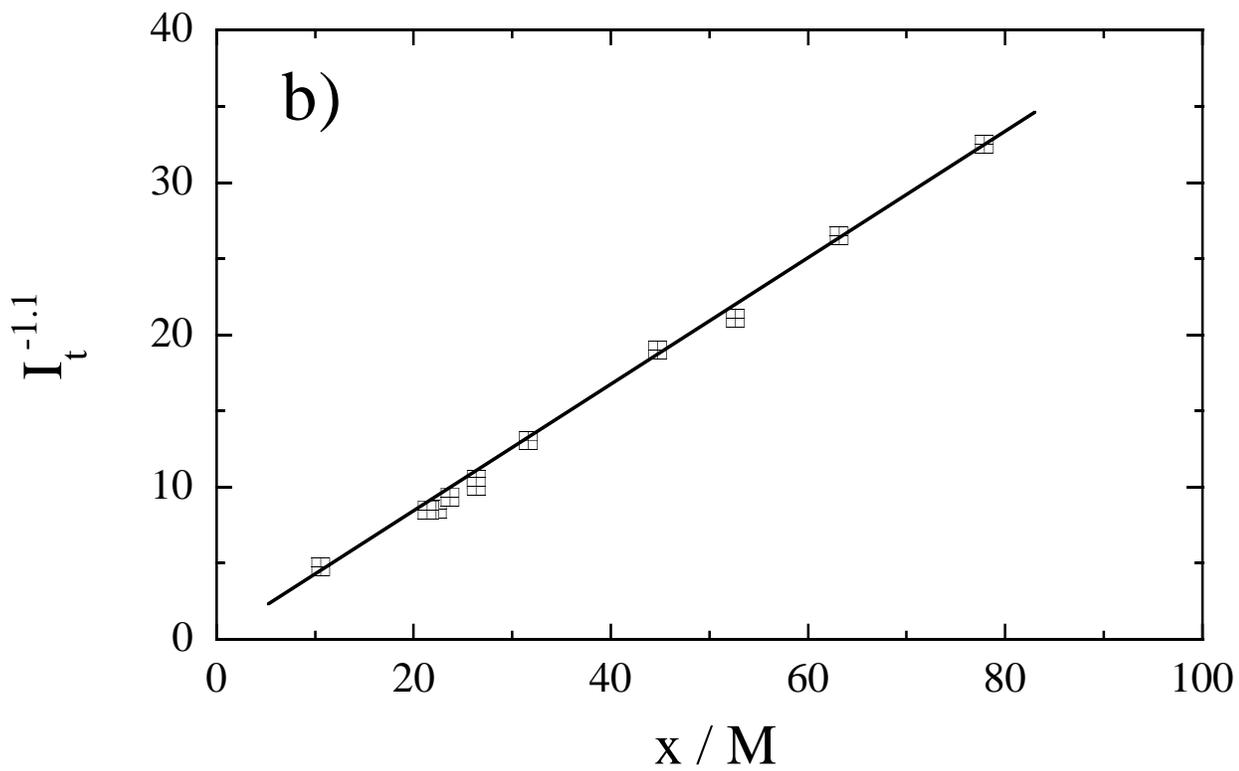
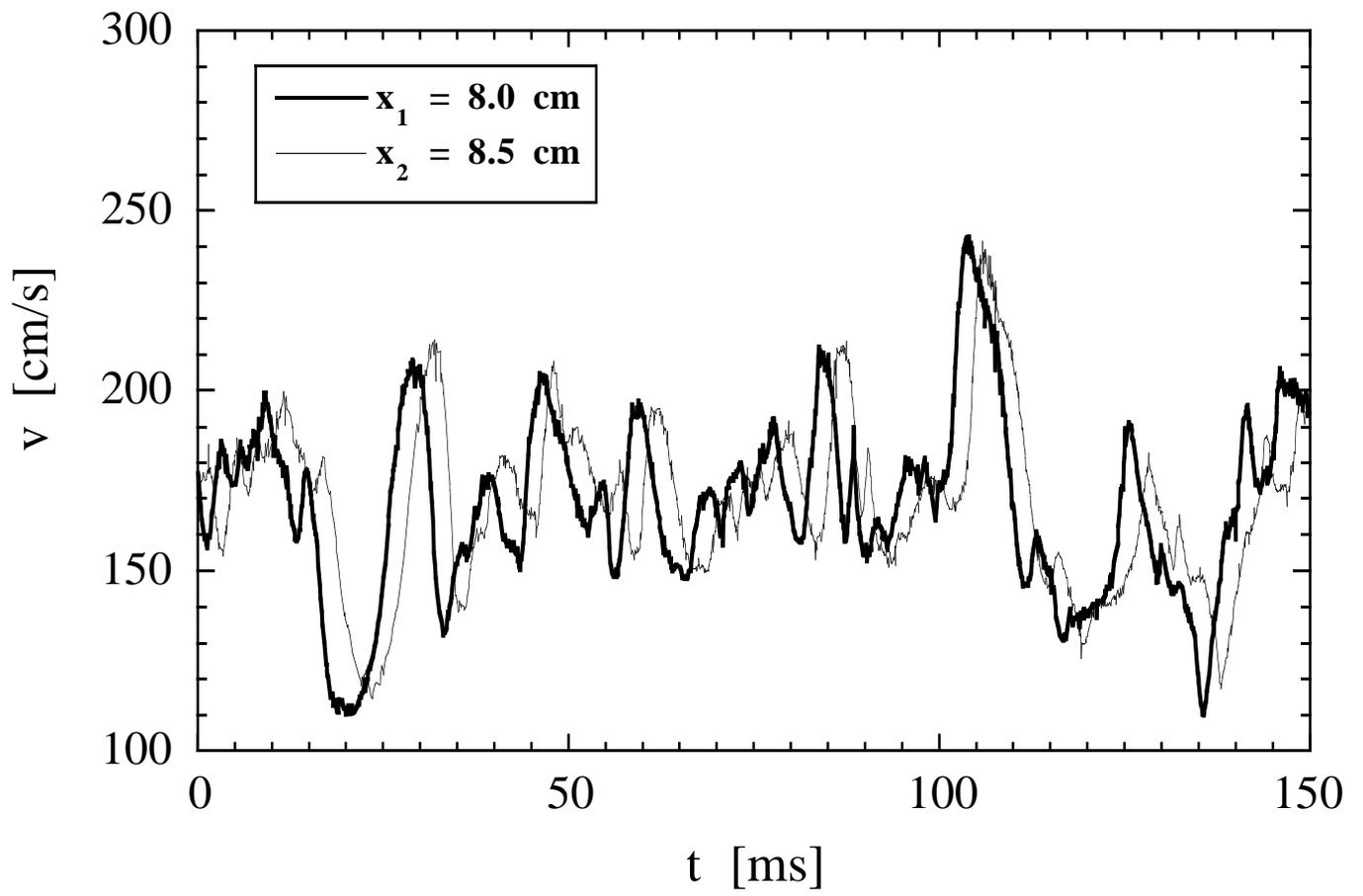
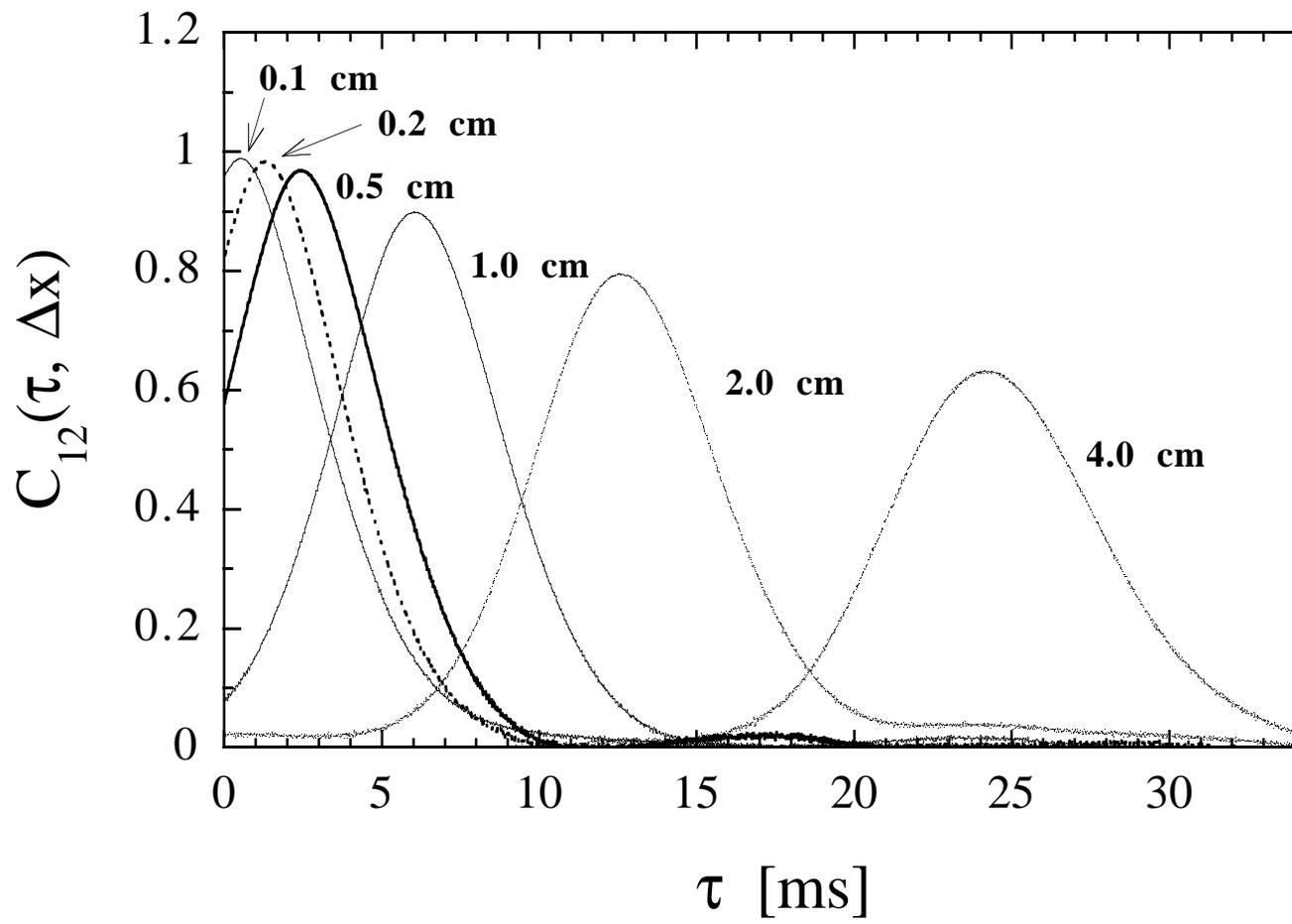
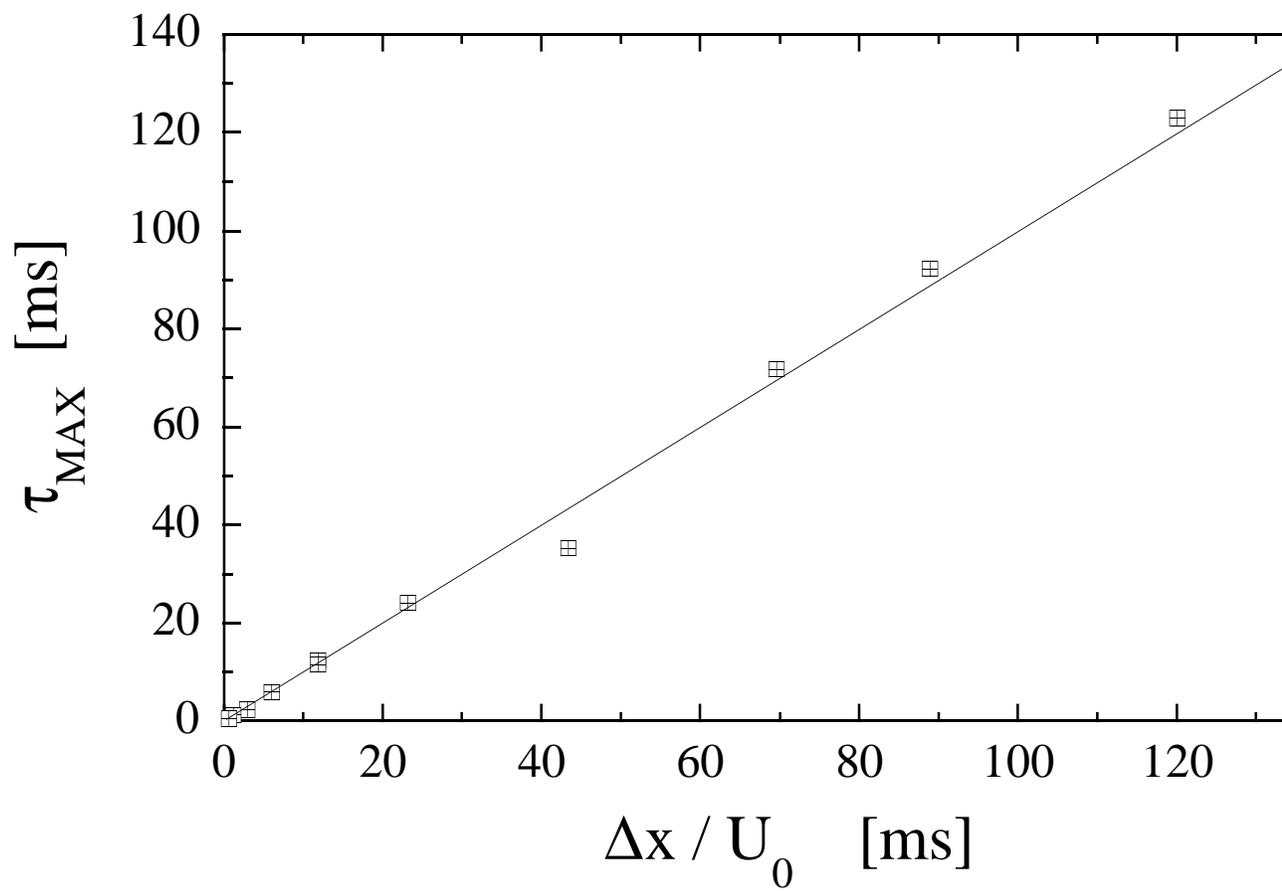
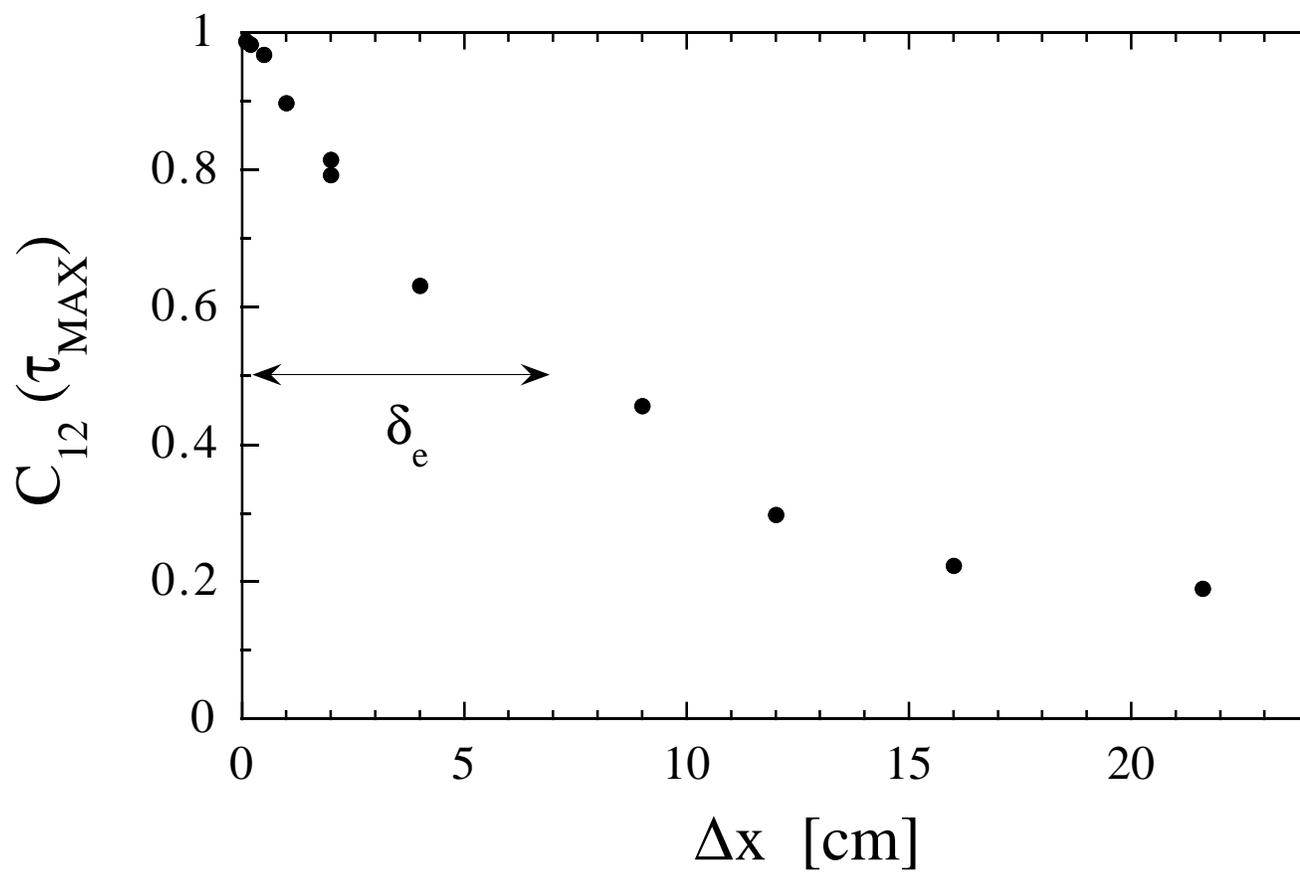
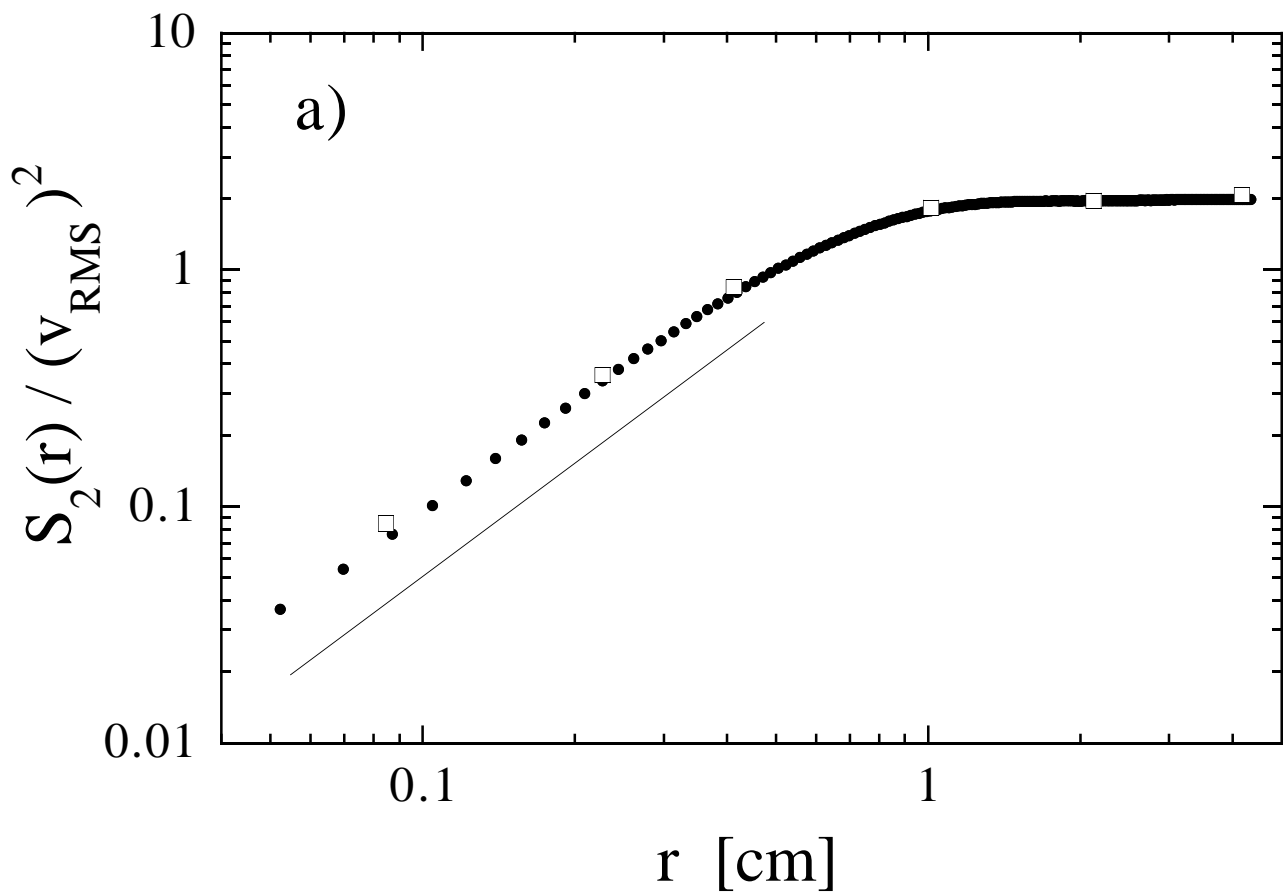
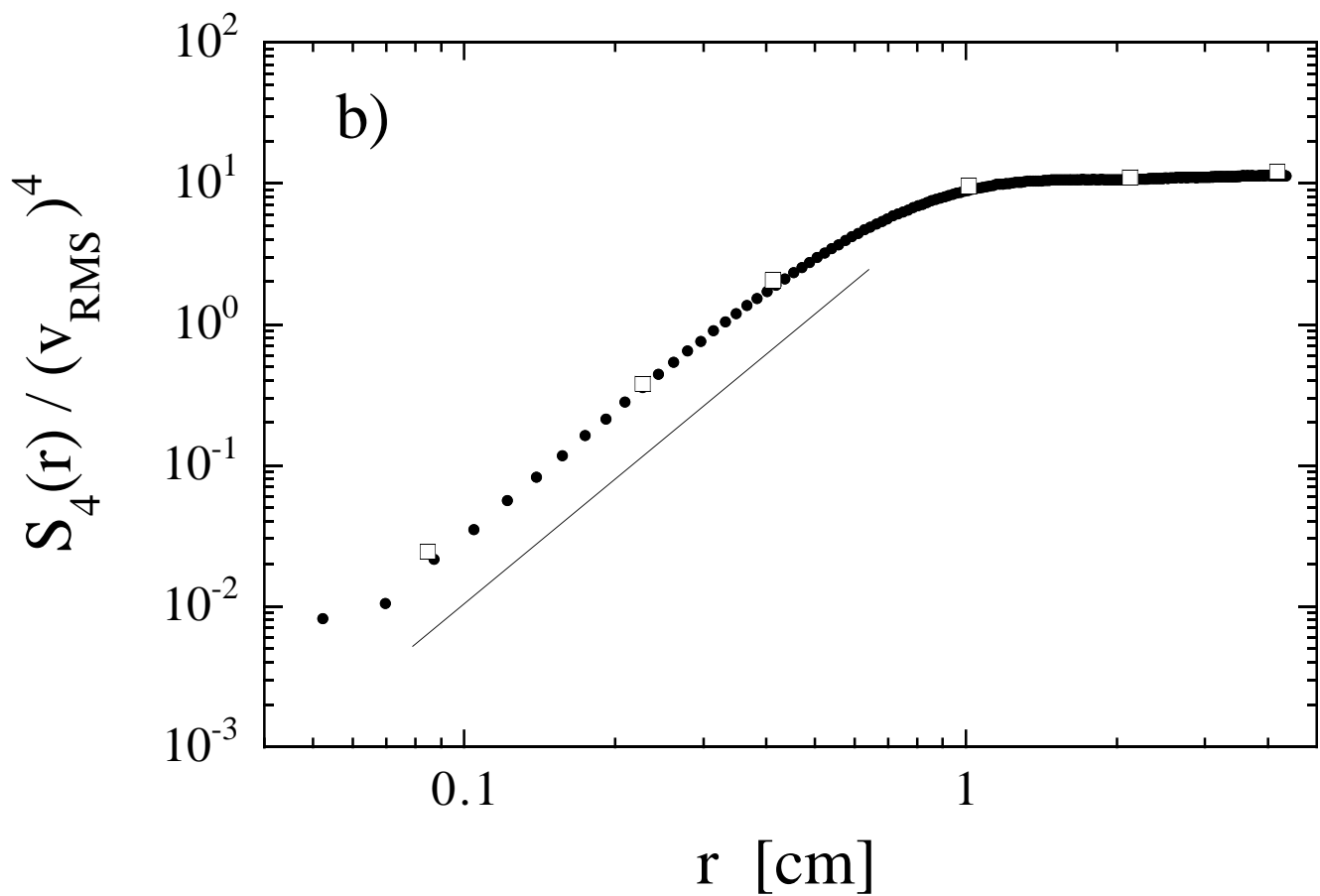
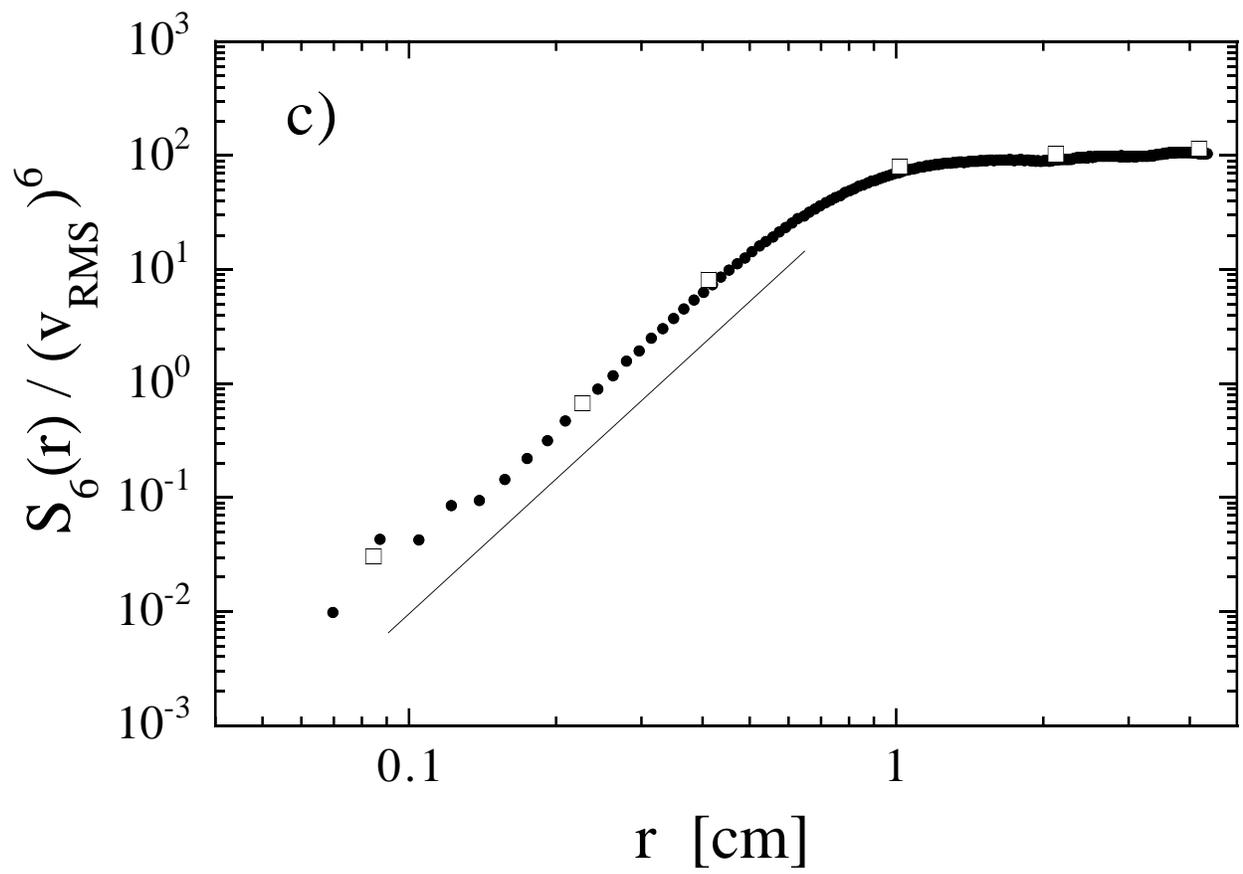
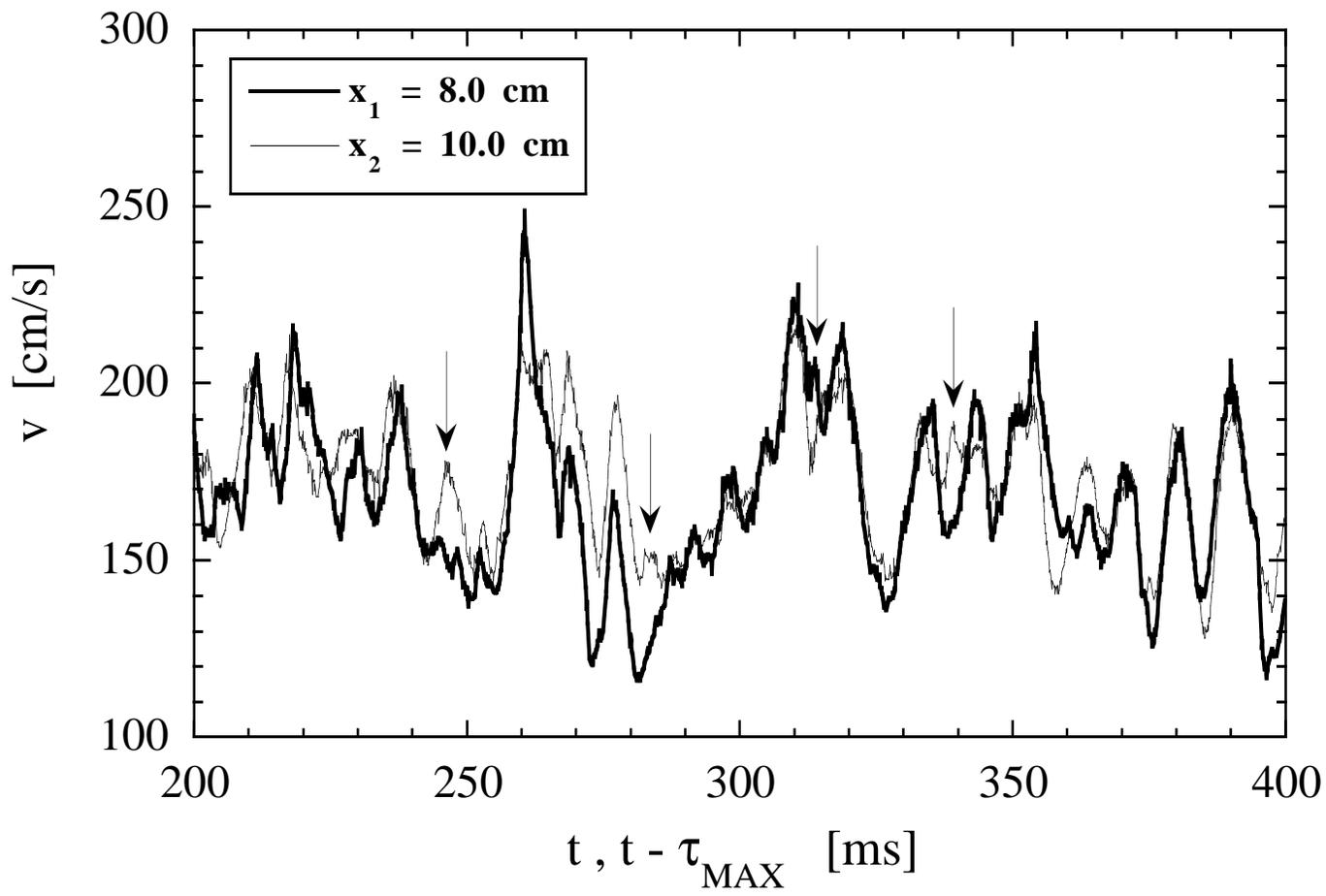
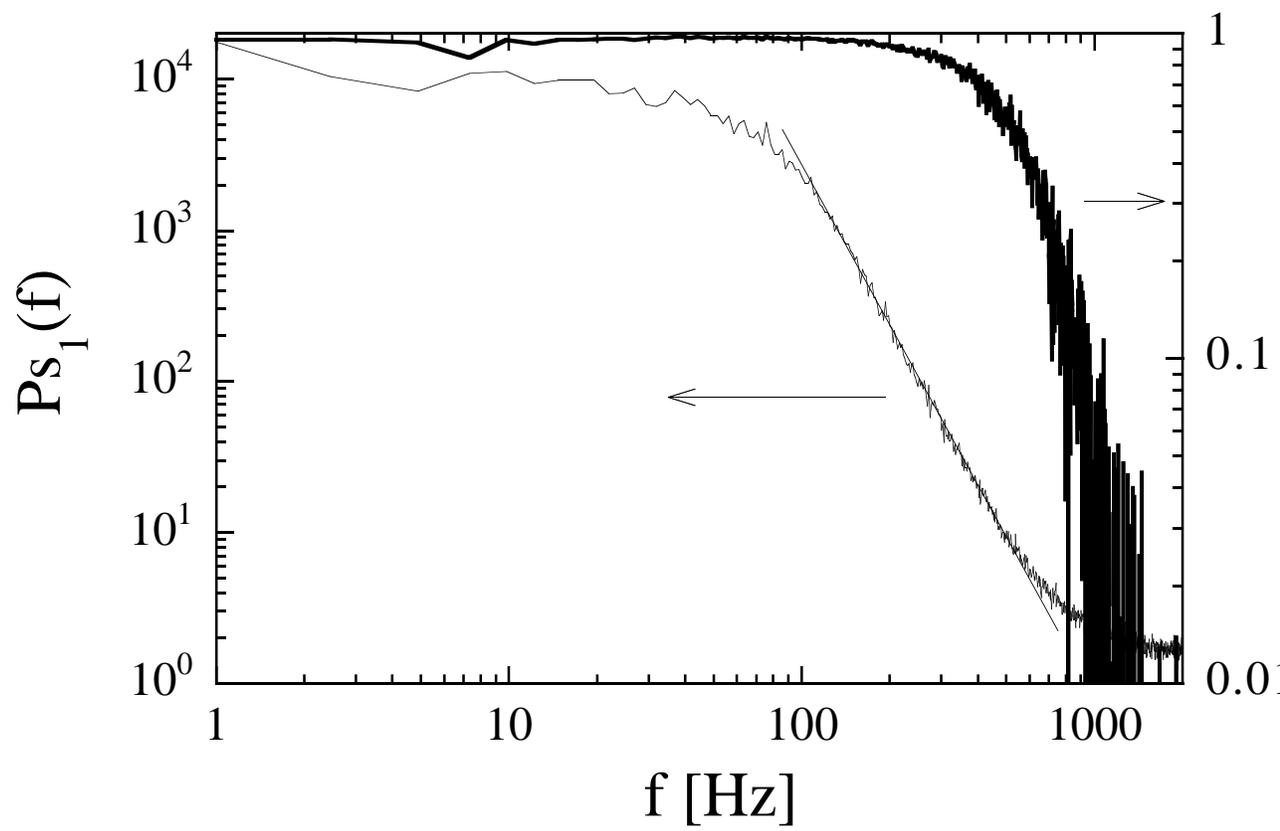
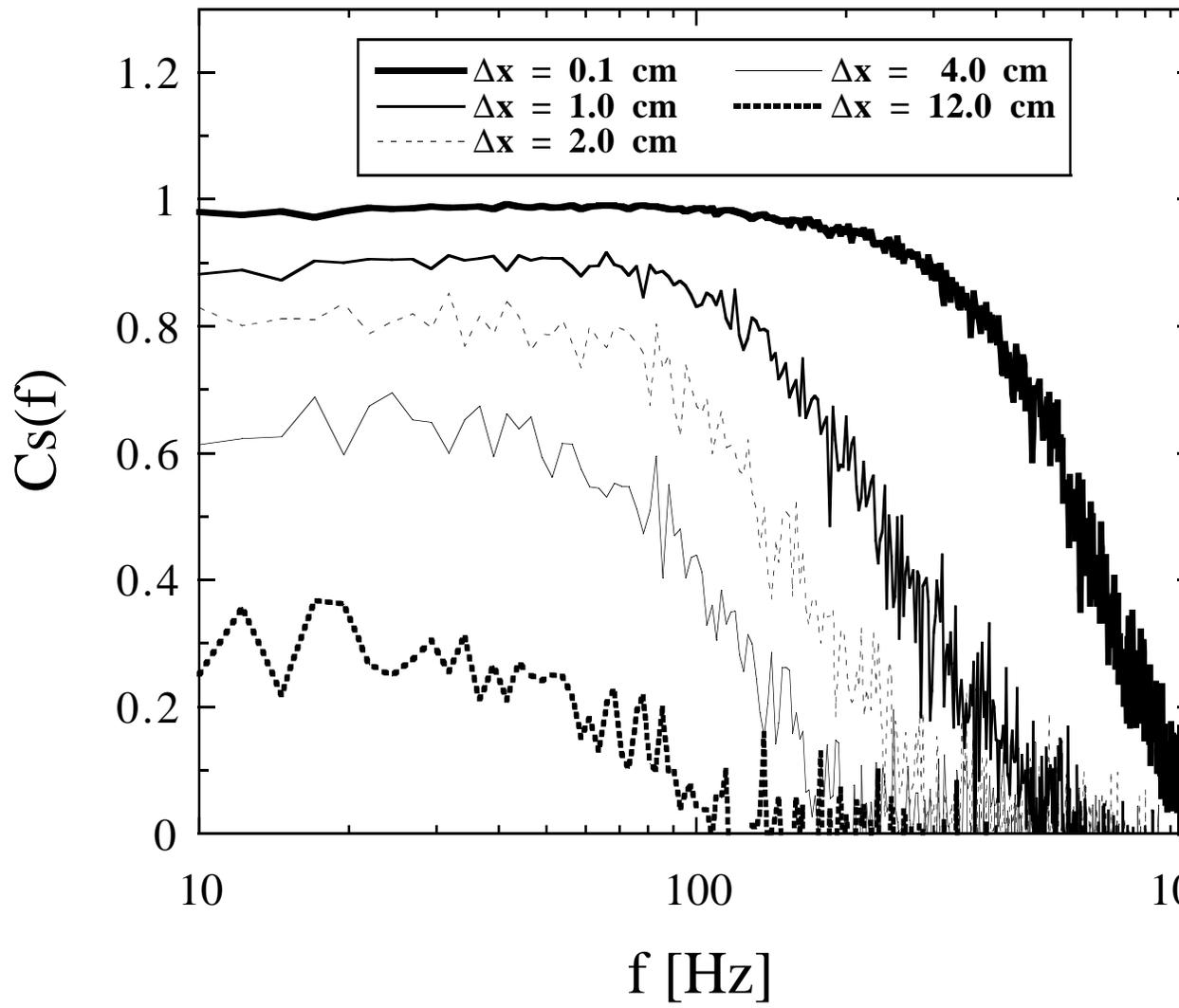
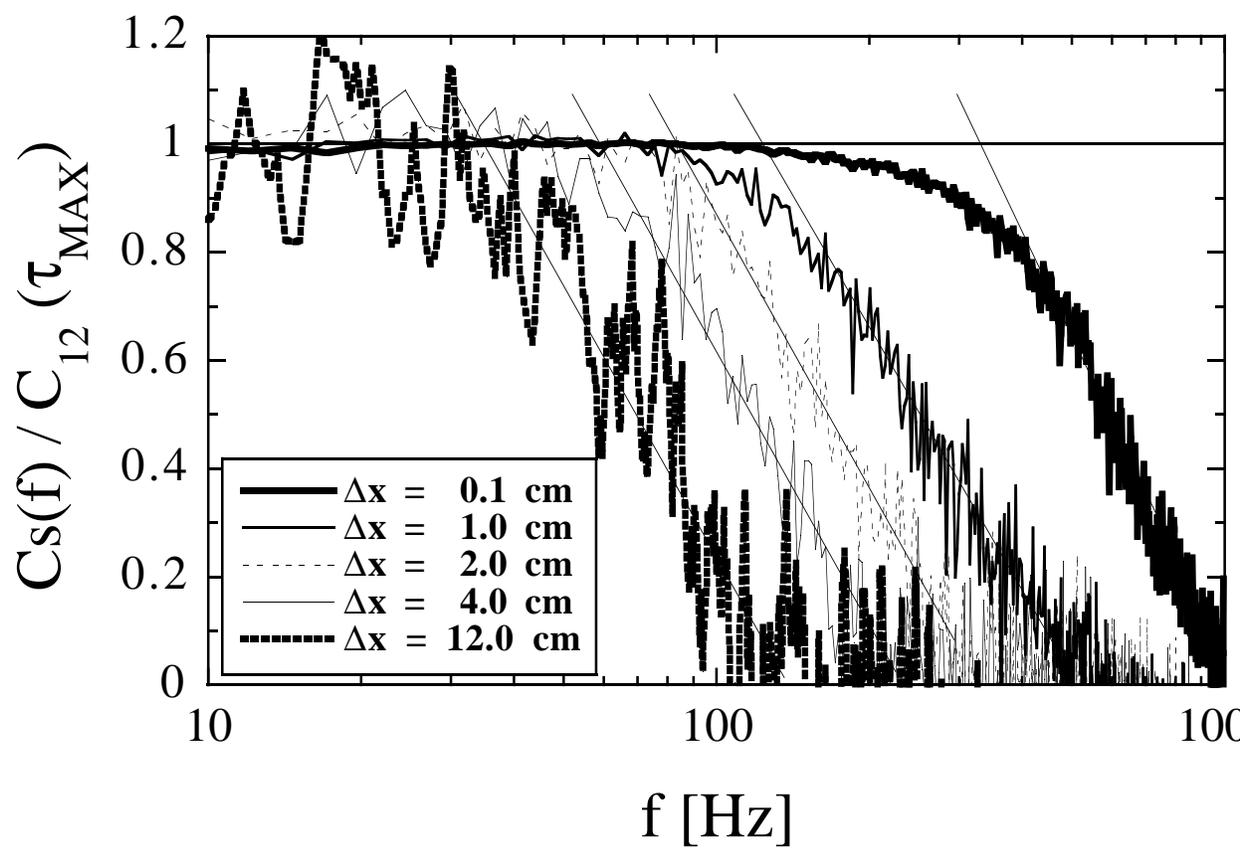
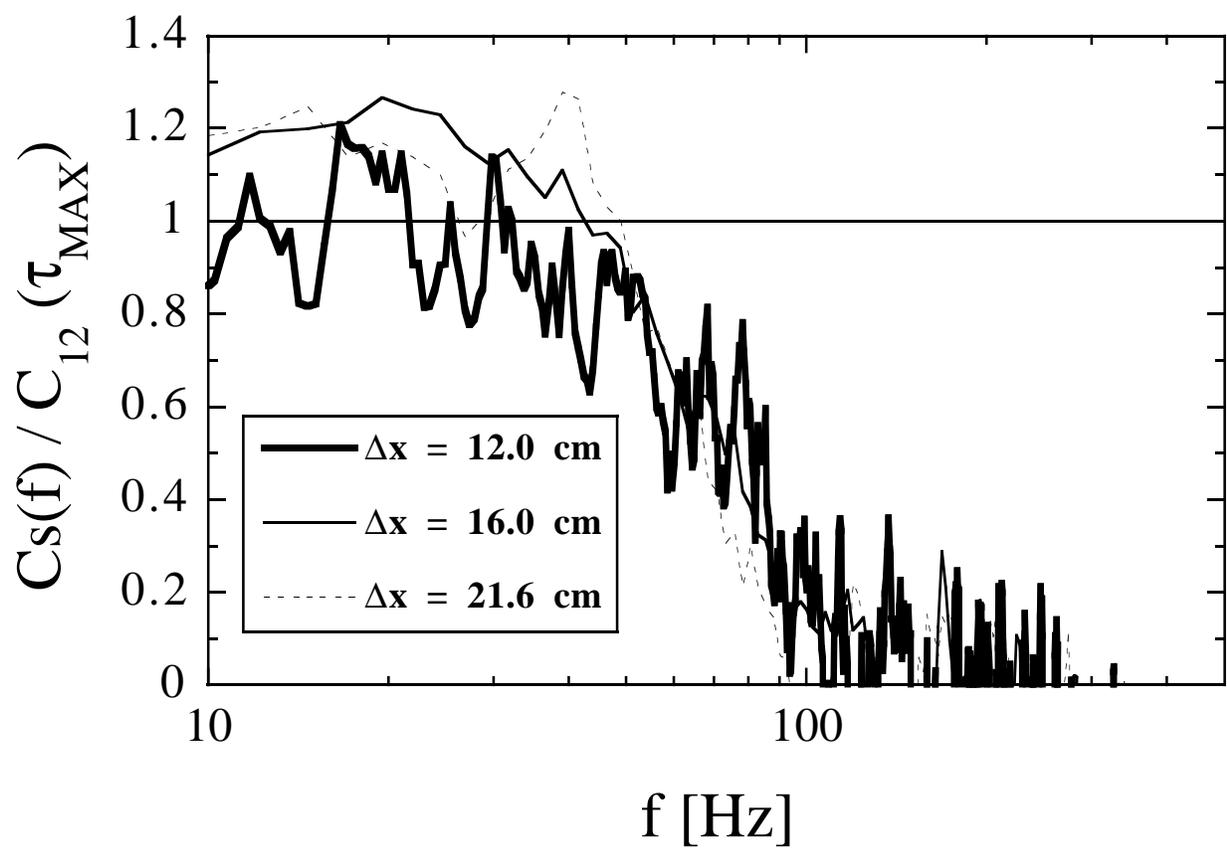
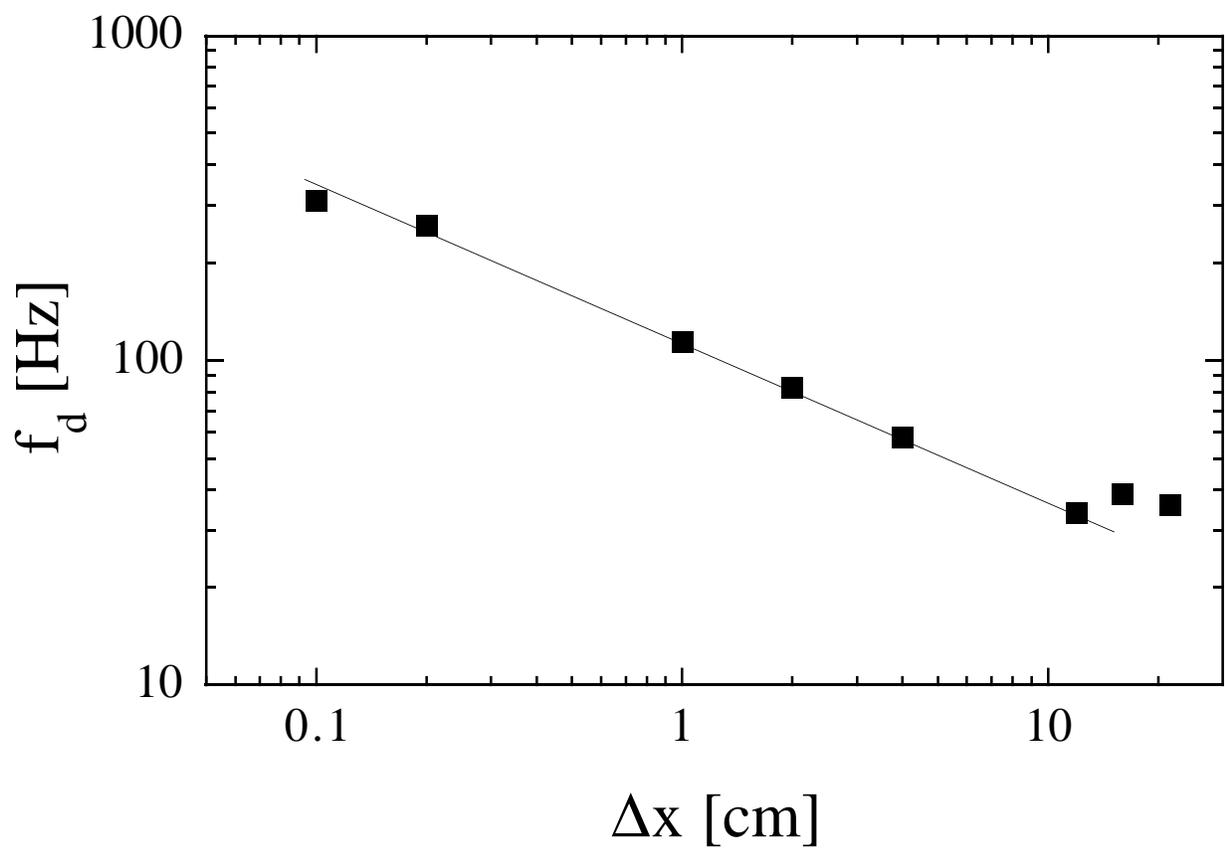
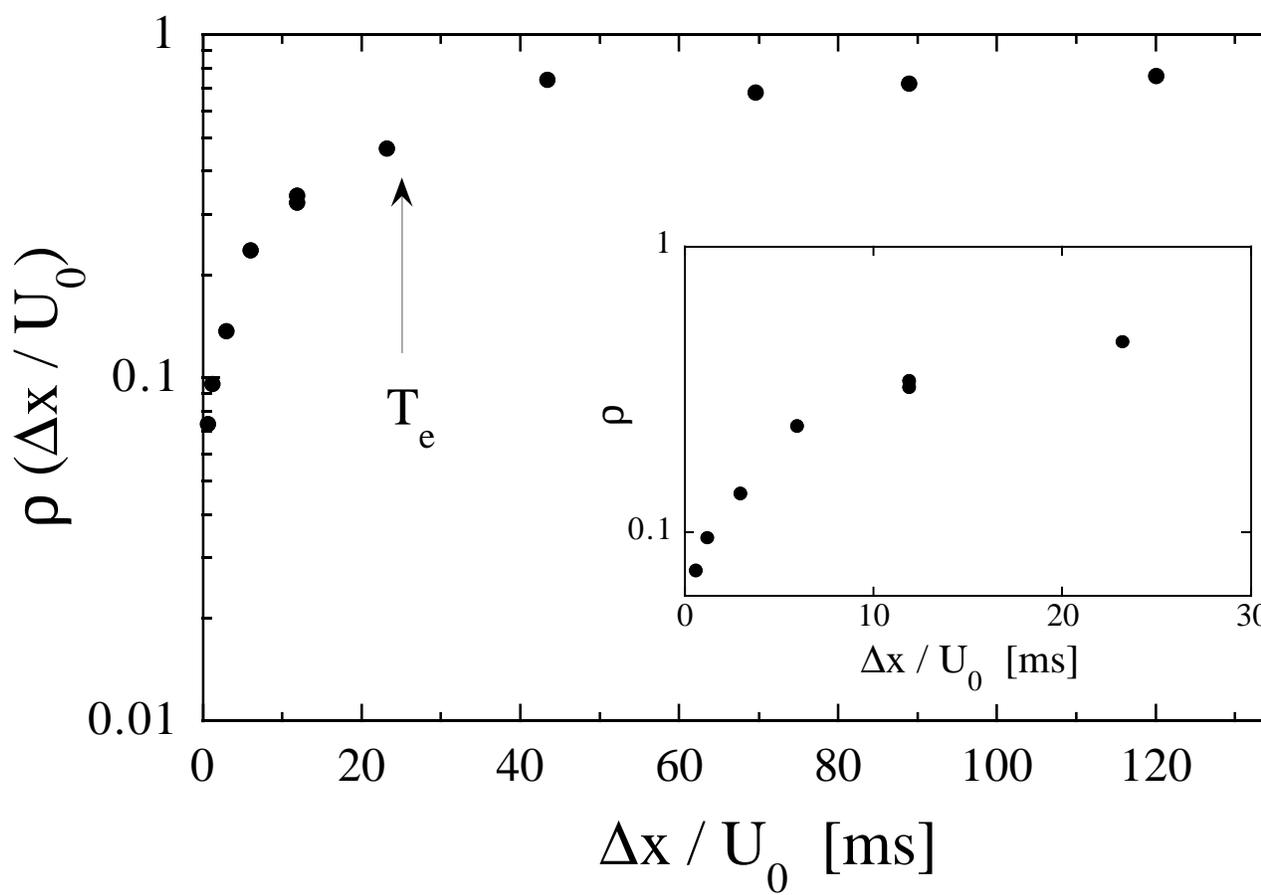
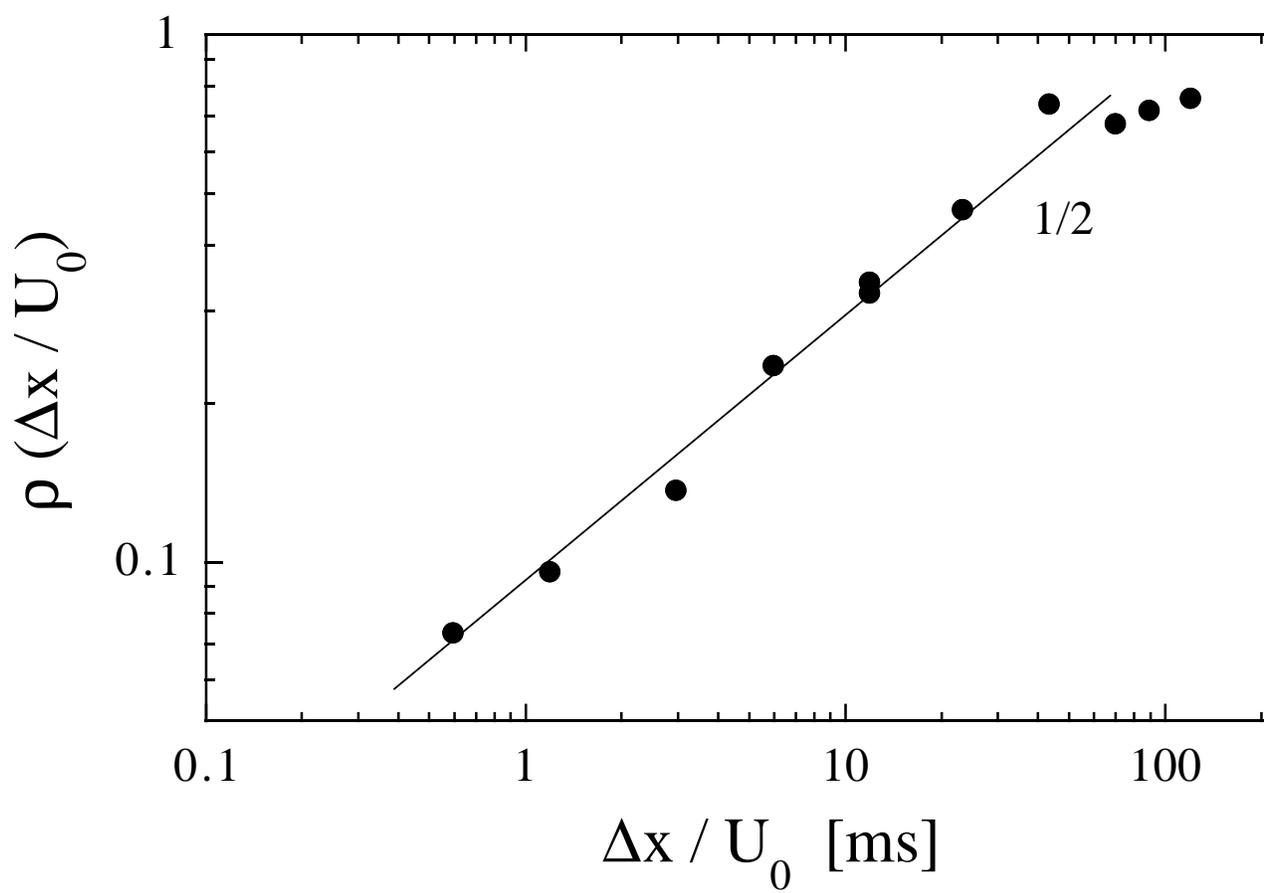
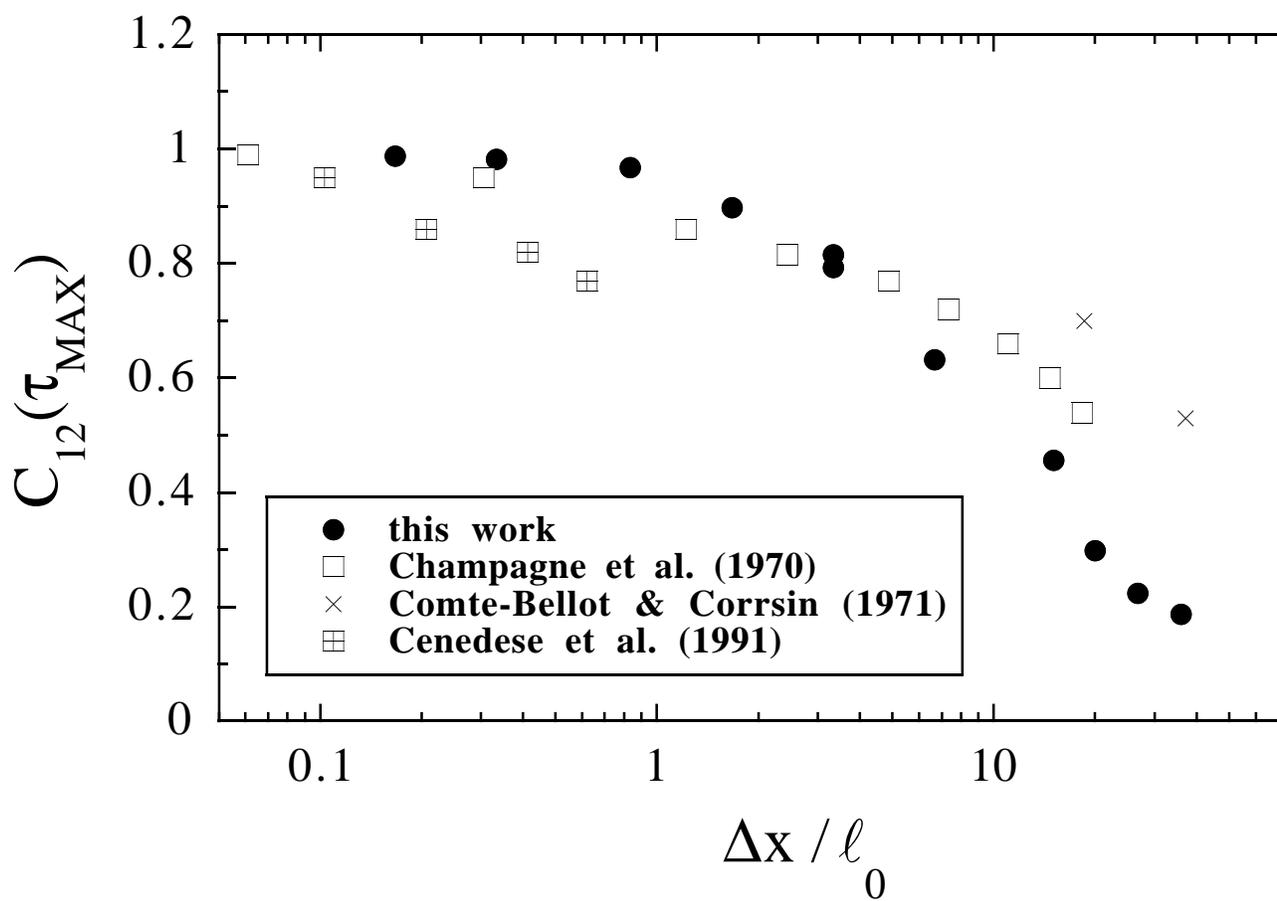
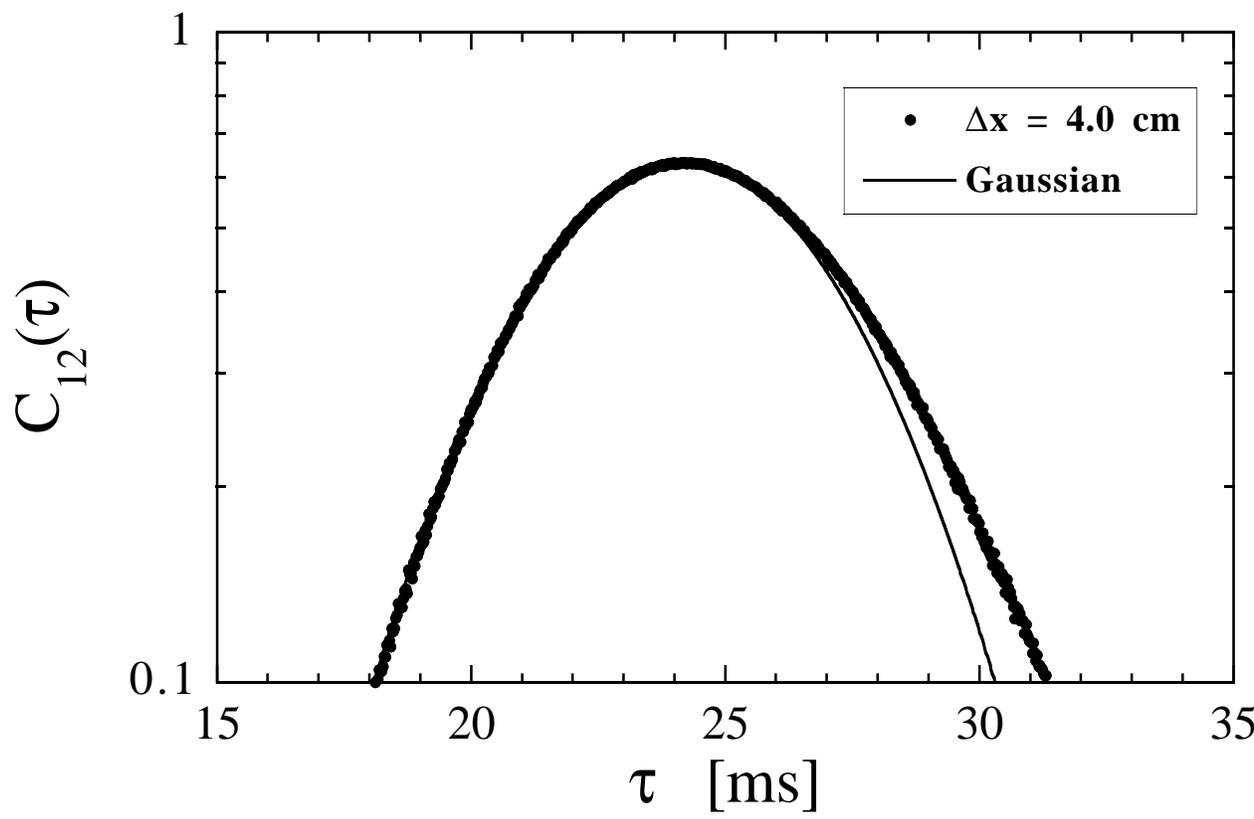
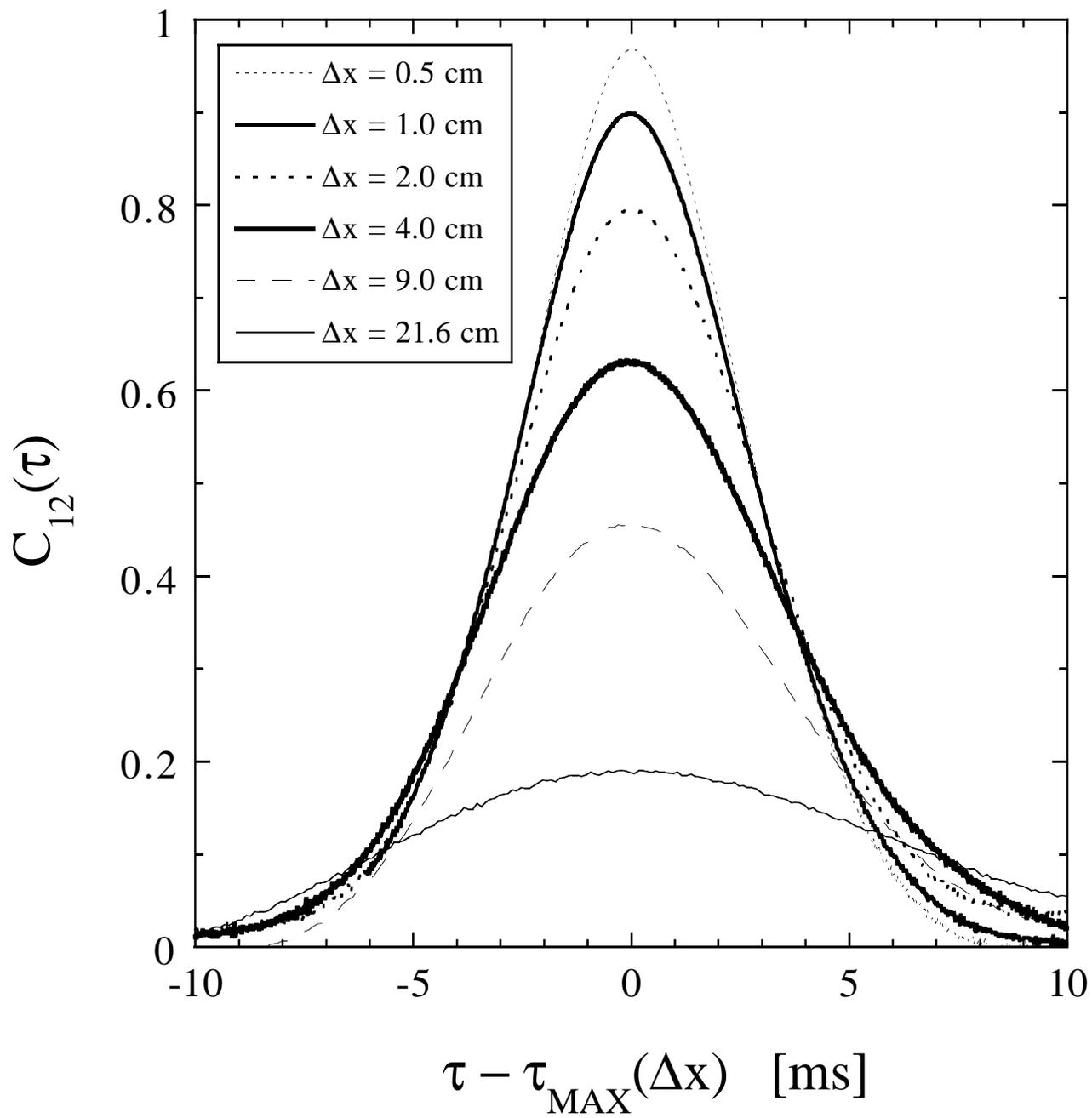
\end{document}